\documentclass[aps,10pt,notitlepage,twocolumn,nofootinbib,superscriptaddress]{revtex4-2}

\usepackage{graphicx,color}
\usepackage{amsmath}
\usepackage{amssymb}
\usepackage{hyperref}
\usepackage[utf8]{inputenc}
\usepackage[english]{babel}
\usepackage{epsfig}
\usepackage{subfigure}
\usepackage{wasysym}
\usepackage{color,xcolor}
\usepackage{amsmath}
\usepackage{bm}
\usepackage{epsfig}
\usepackage{amsfonts}
\usepackage{dcolumn}
\usepackage{float}

\hypersetup{colorlinks=true, linkcolor=blue, citecolor=green}

\usepackage{dcolumn}
\usepackage{bm}
\usepackage{ifpdf}
\usepackage{hyperref}
\usepackage{float}
\usepackage{bm}
\usepackage{xcolor,color,graphicx,graphics}
\usepackage[OT1]{fontenc}
\usepackage{latexsym,amssymb,amsmath,amsfonts}
\usepackage{makeidx}
\usepackage{epsfig}
\usepackage{epstopdf}
\usepackage{mathrsfs}
\hypersetup{colorlinks=true, linkcolor=blue, citecolor=green}
\usepackage{enumerate}
\usepackage{xcolor}
 \usepackage{multirow}

\begin{document}
\title{Black bounces in Cotton gravity}

	\author{Ednaldo L. B. Junior} \email{ednaldobarrosjr@gmail.com}
\affiliation{Faculdade de F\'{i}sica, Universidade Federal do Pará, Campus Universitário de Tucuruí, CEP: 68464-000, Tucuruí, Pará, Brazil}

     \author{José Tarciso S. S. Junior}
    \email{tarcisojunior17@gmail.com}
\affiliation{Faculdade de F\'{i}sica, Programa de P\'{o}s-Gradua\c{c}\~{a}o em F\'{i}sica, Universidade Federal do Par\'{a}, 66075-110, Bel\'{e}m, Par\'{a}, Brazill}

	\author{Francisco S. N. Lobo} \email{fslobo@ciencias.ulisboa.pt}
\affiliation{Instituto de Astrof\'{i}sica e Ci\^{e}ncias do Espa\c{c}o, Faculdade de Ci\^{e}ncias da Universidade de Lisboa, Edifício C8, Campo Grande, P-1749-016 Lisbon, Portugal}
\affiliation{Departamento de F\'{i}sica, Faculdade de Ci\^{e}ncias da Universidade de Lisboa, Edif\'{i}cio C8, Campo Grande, P-1749-016 Lisbon, Portugal}

    \author{\\Manuel E. Rodrigues} \email{esialg@gmail.com}
\affiliation{Faculdade de F\'{i}sica, Programa de P\'{o}s-Gradua\c{c}\~{a}o em F\'{i}sica, Universidade Federal do Par\'{a}, 66075-110, Bel\'{e}m, Par\'{a}, Brazill}
\affiliation{Faculdade de Ci\^{e}ncias Exatas e Tecnologia, Universidade Federal do Par\'{a}, Campus Universit\'{a}rio de Abaetetuba, 68440-000, Abaetetuba, Par\'{a}, Brazil}

 \author{Diego Rubiera-Garcia} \email{ drubiera@ucm.es}
\affiliation{Departamento de Física Téorica and IPARCOS, Universidad Complutense de Madrid, E-28040 Madrid, Spain}

     \author{Luís F. Dias da Silva} 
        \email{fc53497@alunos.fc.ul.pt}
\affiliation{Instituto de Astrof\'{i}sica e Ci\^{e}ncias do Espa\c{c}o, Faculdade de Ci\^{e}ncias da Universidade de Lisboa, Edifício C8, Campo Grande, P-1749-016 Lisbon, Portugal}

    \author{Henrique A. Vieira} \email{henriquefisica2017@gmail.com}
\affiliation{Faculdade de F\'{i}sica, Programa de P\'{o}s-Gradua\c{c}\~{a}o em F\'{i}sica, Universidade Federal do Par\'{a}, 66075-110, Bel\'{e}m, Par\'{a}, Brazill}

\begin{abstract}

Recently, J. Harada proposed a theory relating gravity to the Cotton tensor, dubbed as ``Cotton gravity" (CG). This is an extension of General Relativity such that every solution of the latter turns out to be a solution of the former (but the converse is not true) and, furthermore, it is possible to derive the cosmological constant as an integration constant within it. In this work we investigate CG by coupling it to both non-linear electrodynamics (NLED) and scalar fields.  We study static and spherically symmetric solutions implementing a bouncing behaviour in the radial function so as to avoid the development of singularities, inspired by the Simpson-Visser black bounce and the Bardeen model, both interpreted as magnetic monopoles. We identify the  NLED Lagrangian density and  the scalar field potential generating such solutions, and investigate the corresponding gravitational configurations in terms of horizons, behaviour of the metric functions, and regularity of the Kretchsman curvature scalar. Our analysis extends the class of non-singular geometries found in the literature and paves the ground for further analysis of black holes in CG.

\end{abstract}
\date{\today}
\maketitle
\def\HMS{{\scriptscriptstyle{\rm HMS}}}

\section{Introduction}\label{sec1}

Recently, Harada \cite{Harada:2021bte} proposed a theory that generalizes General Relativity (GR), known as ``Cotton gravity" (CG). It is based on the use of the Cotton tensor, assumed to encapsulate the effects describing gravity, and leading to third-order field equations in the derivatives of the metric tensor  \cite{Harada}. 
Indeed, the Cotton tensor \cite{Cottonpaper} is a rank-3 tensor that captures the curvature properties of Riemannian manifolds in $n$ dimensions. It vanishes for manifolds with $n < 3$ and only appears in three-dimensional spaces where it becomes zero if, and only if, the manifold is conformally flat. Historically, the Cotton tensor has been primarily used to study three-dimensional manifolds \cite{Cottonpaper} and gravity in lower dimensions \cite{ Jackiw:2004qm, Deser:2004wd}. However, as discussed in \cite{Harada:2021bte}, it has been discovered that this tensor also plays a crucial role in describing gravitational effects beyond GR in four-dimensional spacetime. The Cotton tensor's emergence provides a different approach to describing spacetime curvature, emphasizing the conformal properties of gravity, which GR does not inherently address.
Part of the interest in this theory also comes from the fact that  every solution of  Einstein's equations of GR (including a cosmological constant term) is a solution of CG gravity via a particular Cotton tensor. Indeed, the cosmological constant itself arises as an integration constant of the theory. Furthermore, in \cite{Harada:2022edl} it was shown that a particular new solution of CG gravity not present within GR could be able to explain the galactic rotation curves without the need for dark matter, further raising the interest in the theory. As a consequence, different applications of this theory have been studied very recently in the literature \cite{Sussman:2023eep,Mantica:2022flg,Sussman:2023wiw,Gogberashvili:2023wed,Mantica:2023ssd,Xia:2024tps,Mo:2024rfq,Junior:2023ixh,Junior:2024vrv,Mantica:2023stl,Harada:2023afu,Barnes:2023uru,Barnes:2023qfi,Barnes:2024vjq,Mantica:2024mun}. 

More specifically, a key feature of CG is its ability to explain the accelerating expansion of the universe without invoking dark energy. In GR, this acceleration is typically attributed to a cosmological constant or dark energy. However, in CG, the cosmological constant can be interpreted as a manifestation of spatial curvature, offering a geometrically motivated alternative to dark energy \cite{Harada,Sussman:2023wiw}. Thus, CG provides an elegant geometric alternative to account for cosmic phenomena.
Recent work in CG has yielded a variety of new exact solutions, including spherically symmetric spacetimes and black hole solutions. The authors in \cite{Gogberashvili:2023wed,Junior:2023ixh,Junior:2024vrv} explore how CG introduces novel black hole solutions that differ from those predicted by GR. These solutions could potentially provide new insights into the nature of black holes and gravitational waves, indicating the theoretical flexibility of CG compared to the more constrained structure of GR.
Furthermore, CG, particularly in its extension through conformal Killing gravity, offers deeper insights into the structure of spacetime and provides a rich framework for understanding the role of conformal symmetries in cosmology and gravity \cite{Harada:2023afu,Mantica:2024mun}. The study of conformal Killing vector fields introduces new approaches to understanding the evolution of the universe, especially regarding the growth of cosmic structures and the interaction of gravity with the dark sector, providing a broader set of tools than GR \cite{Mantica:2022flg,Mantica:2023ssd}.

In fact, there is a current debate on the theory as a viable extension of GR \cite{Clement:2023tyx,Sussman:2024iwk,Clement:2024pjl,Sussman:2024qsg}, which highlights the need for further work to determine the potential of Cotton gravity to supersede GR.
In particular, the concern about whether CG can remain fully compatible with the equivalence principle, which is a cornerstone of GR, has been a key part of the debate. In \cite{Clement:2023tyx}, the authors suggest that modifications introduced by CG might violate this principle. They argue that the generalizations inherent in CG could lead to situations where local inertial frames are no longer indistinguishable from free-falling frames, as required by the equivalence principle.
However, in \cite{Sussman:2024iwk,Sussman:2024qsg}, the authors address this concern by clarifying that CG still respects a generalized form of the equivalence principle within its framework. They argue that the Cotton tensor, which is central to their formulation, adds new geometric structures without undermining the core symmetry principles that satisfy the equivalence principle. In particular, they demonstrate that CG retains consistency with key spacetime symmetries and, when applied to appropriate solutions, does not inherently violate the equivalence principle. They maintain that the critique overlooks the specific conditions under which the theory was constructed, and that these conditions still allow for the equivalence principle to hold.

The second major concern is whether CG can withstand observational scrutiny, especially in light of data from gravitational wave detectors like LIGO/Virgo and black hole observations. In \cite{Clement:2023tyx,Clement:2024pjl}, the authors express skepticism about the theory's ability to predict phenomena consistent with current observational data. They argue that the mathematical structures introduced by CG may lead to results that diverge from well-established predictions of GR, particularly concerning gravitational wave signals and black hole dynamics.
In response, in \cite{Sussman:2024iwk,Sussman:2024qsg} the authors acknowledge that CG, as a theoretical framework, is still in its developmental phase in terms of making precise observational predictions. However, they argue that dismissing the theory based on its current lack of direct observational support is premature. In fact, CG introduces new mathematical insights into spacetime curvature, and testing these ideas against observational data requires further refinement of the theory’s specific solutions. The authors of \cite{Sussman:2024iwk,Sussman:2024qsg} point out that while CG may not produce the same gravitational wave signals or black hole behaviors as GR, this does not necessarily invalidate the theory. Instead, it highlights the need for more work to derive concrete, testable predictions within this new framework.
Moreover, they stress that CG's potential to provide new insights into gravitational phenomena, particularly in regions of extreme curvature (such as near black hole singularities), makes it worthy of further exploration. They argue that the failure to immediately match observations does not equate to failure of the theory, as observational tests for any modified gravity theory require careful and specific theoretical modelling.

One of the central points raised in \cite{Clement:2023tyx,Clement:2024pjl} is that CG lacks the predictive power necessary to be considered a viable alternative to GR. The authors of \cite{Clement:2023tyx,Clement:2024pjl} claim that, given the highly constrained nature of observational tests, CG must either match the predictions of GR in key regimes or propose novel phenomena that can be observed to validate the theory.
In their replies, the authors of \cite{Sussman:2024iwk,Sussman:2024qsg} counter this by arguing that CG introduces novel mathematical structures that have yet to be fully explored in terms of their observational implications. While it may be true that the theory does not yet provide a fully detailed set of testable predictions for gravitational waves or black hole dynamics, the authors defend CG as a significant extension of the geometric foundations of gravity. They argue that new theories often face the challenge of being tested against observations only after significant theoretical groundwork has been laid. Indeed, CG is in the phase of this groundwork being developed, and it is too early to dismiss the theory on the grounds that it is not immediately predictive in the same way as GR.
Additionally, they argue that the theory's ultimate value lies in its ability to generalize Einstein's field equations in a way that could reveal new insights into spacetime geometry. This opens up the possibility that, as the theory is further developed and tested, it may indeed lead to predictions that could be tested with future gravitational wave or black hole observations.

Thus, the concerns raised about CG, specifically regarding the equivalence principle and observational constraints, are valid and significant, one may argue that these concerns do not represent fatal flaws in the theory but rather areas for further development and testing. CG, as a novel theoretical framework, respects a generalized form of the equivalence principle and introduces new geometrical structures that can still be consistent with known physics. While it has yet to produce concrete observational predictions, the theory's potential contributions to understanding the geometry of spacetime suggest that further research is necessary before definitive conclusions can be drawn.

Among the many gravitational configurations explored in the literature beyond the canonical (Kerr) black hole of GR, those aimed to solve the singularity problem are of particular interest. Such a problem is related to the unavoidably development of space-time singularities (as given by the incompleteness of at least one geodesic trajectory) inside black holes within the GR framework (see \cite{Senovilla:2014gza} for a discussion of this topic), and which typically correlates with the presence of infinities in some sets of curvature scalars. A canonical strategy in the field at this regard is to imbue the radial function with a non-trivial bouncing behaviour in order to prevent the focusing of geodesics appearing in the singularity theorems, this way allowing geodesic congruences to re-expand at a minimum radial distance. Examples of this are the well known Bardeen model \cite{Bardeen,Ayon-Beato:2000mjt} and, more recently, the class of black bounces introduced by Simpson and Visser in \cite{Simpson:2018tsi}. Typically, models of this kind introduce one (or more) parameters associated to a minimum length scale, which is responsible for the bouncing behaviour. Furthermore, the bounce can be hidden behind event horizons thus representing regular black holes, or be naked instead, interpreted as a kind of traversable wormhole allowing free communication between far-away distances \cite{Lobo:2020kxn}. 

The various forms of the black bounce proposal have led to the development of several additional classes of solutions~\cite{Junior:2022zxo,Junior:2023qaq,Junior:2024xmm,Rodrigues:2022mdm,Huang:2019arj} and the analysis of their properties. Among them, we can underline regularity, causal structure and energy conditions \cite{Lobo:2020ffi}, gravitational lensing  \cite{Nascimento:2020ime, Tsukamoto:2020bjm, Cheng:2021hoc, Tsukamoto:2021caq, Zhang:2022nnj}, and observational  signatures \cite{Guerrero:2021ues,Jafarzade:2021umv,Jafarzade:2020ova,Jafarzade:2020ilt,Yang:2021cvh,Bambhaniya:2021ugr,Ou:2021efv,Guo:2021wid,Wu:2022eiv,Tsukamoto:2022vkt}, besides generalizations to the rotating case \cite{Mazza:2021rgq,Xu:2021lff}. Typically, such bouncing configurations are supported by a combination of a non-linear electrodynamics (NED) and a scalar field  ~\cite{Bronnikov:2022bud,Canate:2022gpy,Rodrigues2023,Pereira:2023lck} working under a reverse procedure: the geometrical line element incorporating the bouncing behaviour is set first, and then the field equations are driven back to find the explicit matter sources treading the geometry.

The main aim of this work is to implement suitable generalizations of both Simpson-Visser and Bardeen solutions within CG gravity according to the new features of the theory. In particular, we shall generalize the exact vacuum solution of CG of \cite{Harada:2021bte}, characterized by a new parameter $\gamma$, towards bouncing geometries supported by NLED and scalar fields. The geometries obtained this way depend on the mass of the black hole $M$, on the magnetic charge $q$, and on the CG parameter $\gamma$, besides the cosmological constant itself $\Lambda$. Given the four-parametric nature of these CG solutions, we shall take different choices for them in both the Simpson-Visser and Bardeen solutions, hopefully covering the most relevant conceptual classes of configurations in terms of types of horizons and global shape of the corresponding metric functions. We shall then obtain the functional shapes of the NLED Lagrangian density and the scalar field potential.


This work is organized as follows. In Sec. \ref{sec2}, we briefly present the field equations of CG coupled to NLED and scalar fields, introduce the class of geometries we are interested in, and find the equations allowing to reconstruct the matter (NLED supported by magnetic charge) and scalar field potential. In Sec. \ref{BB_SV}, we  generalize the black bounce geometries and study in detail their properties, and in Sec. \ref{BB_Bardeen} we do the same with the Bardeen one. In Sec.  \ref{sec:concl}, we summarize and discuss our results. We work in units $G=c=1$.

\section{Cotton Gravity coupled to non-linear electrodynamics and scalar fields}\label{sec2}

\subsection{Field equations of Cotton gravity}

In CG the gravitational field equations are given by  \cite{Harada:2021bte} \begin{equation}
C_{\alpha\mu\nu}=\kappa^2\nabla_\beta\Theta^\beta_{\phantom{\beta}\alpha\mu\nu},\label{eq_CG}
\end{equation}
where $\kappa^2=8\pi$, while the tensors $C_{\alpha\mu\nu}$ and $\Theta_{\alpha\mu\nu}$ are defined as
\begin{eqnarray}
C_{\alpha\mu\nu}&\equiv&\nabla_{\mu}R_{\alpha\nu}-\nabla_{\nu}R_{\alpha\mu}
\nonumber\\
&&-\frac{1}{6}\left(g_{\nu\alpha}\nabla_{\mu}R-g_{\mu\alpha}\nabla_{\nu}R\right),
	 \\
\nabla_{\beta}\Theta_{\phantom{\mu}\alpha\mu\nu}^{\beta}&\equiv&\frac{1}{2}\left(\nabla_{\mu}\Theta_{\alpha\nu}-\nabla_{\nu}\Theta_{\alpha\mu}\right)\nonumber\\
&&-\frac{1}{6}\left(g_{\nu\alpha}\nabla_{\mu}-g_{\mu\alpha}\nabla_{\nu}\right)\Theta \,.
\end{eqnarray}
respectively, where $C_{\mu\nu}$ stands for the Cotton tensor, $R_{\mu\nu}$ for the Ricci tensor and $\Theta_{\mu\nu}$  for the energy-momentum tensor, and their traces are represented by $R$ and $\Theta$, respectively. Moreover, Einstein's equations of GR are solutions of Eq.~\eqref{eq_CG}, and therefore, GR can be seen as a particular case of CG. Note that if we contract the equation of motion ~\eqref{eq_CG} with $g^{\alpha\nu}$, the conservation law $g^{\alpha\nu}C_{\alpha\mu\nu}=\kappa^{2}\nabla_{\beta}T_{\phantom{\beta}\mu}^{\beta}=0$ is naturally  satisfied.

For the matter sector of the theory, in this work we are interested in space-times sourced by NLEDs and scalar fields, whose energy-momentum tensors can be written as
\begin{equation}
\nabla_{\beta}\Theta_{\phantom{\mu}\alpha\mu\nu}^{\beta}=\nabla_{\beta}\overset{F}{\Theta}{}_{\phantom{\mu}\alpha\mu\nu}^{\beta}+\nabla_{\beta}\overset{\varphi}{\Theta}{}_{\phantom{\mu}\alpha\mu\nu}^{\beta}.\label{TEM}
\end{equation}
Next, we will explicitly show the forms of these matter components described by the energy-momentum tensor:
\begin{eqnarray}
\nabla_{\beta}\overset{F}{\Theta}{}_{\phantom{\mu}\alpha\mu\nu}^{\beta}	&\equiv&\frac{1}{2}\left(\nabla_{\mu}\overset{F}{T}_{\alpha\nu}-\nabla_{\nu}\overset{F}{T}_{\alpha\mu}\right)\nonumber\\
	&&-\frac{1}{6}\left(g_{\nu\alpha}\nabla_{\mu}-g_{\mu\alpha}\nabla_{\nu}\right)\overset{F}{T},\\
\nabla_{\beta}\overset{\varphi}{\Theta}{}_{\phantom{\mu}\alpha\mu\nu}^{\beta}	&\equiv&\frac{1}{2}\left(\nabla_{\mu}\overset{\varphi}{T}_{\alpha\nu}-\nabla_{\nu}\overset{\varphi}{T}_{\alpha\mu}\right)\nonumber\\
&&	-\frac{1}{6}\left(g_{\nu\alpha}\nabla_{\mu}-g_{\mu\alpha}\nabla_{\nu}\right)\overset{\varphi}{T}\,,
\end{eqnarray}
respectively, with
\begin{align}
    \overset{F}{T}{}_{\mu\nu}&=g_{\mu\nu}{\cal L}_{\rm NLED}(F)-{\cal L}_{F}F_{\mu\alpha}F_{\nu}^{\phantom{\nu}\alpha},\\
\overset{F}{T}&=4{\cal L}_{\rm NLED}(F)-4{\cal L}_{F}F,\\
    \overset{\varphi}{T}{}_{\mu\nu}&=2\,\epsilon\,\partial_{\mu}\varphi\partial_{\nu}\varphi-\epsilon 
g_{\mu\nu}\partial^{\sigma}\varphi\partial_{\sigma}\varphi+g_{\mu\nu}V(\varphi),\\
\overset{\varphi}{T}&=-2\,\epsilon\,\partial^{\nu}\varphi\partial_{\nu}\varphi+4V(\varphi),
\end{align}
where  $\epsilon=+1$ is the scalar field, while $\epsilon=-1$ represents the phantom scalar field, $\varphi$ is the scalar field, while $V(\varphi)$ denotes the scalar potential,  ${\cal L}_{\rm NLED}(F)$ is the Lagrangian density describing a NLED that depends on the electromagnetic scalar
\begin{equation}
F=\frac{1}{4}F^{\mu\nu}F_{\mu\nu},    \label{F}
\end{equation}
 and the  Maxwell-Faraday antisymmetric tensor is defined by $ F_{\mu \nu} = \partial_\mu A_\nu -\partial_\nu A_\mu$,
 where ${A_\alpha}$ is the vector potential. 
 
 The field equations for these two matter contributions can be found by the usual variational procedure as 
\begin{align}
\nabla_\mu ({\cal L}_F F^{\mu\nu})&=\frac{1}{\sqrt{-g}} \partial_\mu (\sqrt{-g} {\cal L}_F F^{\mu\nu})=0,\label{sol2}\\
     2\nabla_\nu\nabla^\mu \varphi&=-\frac{dV(\varphi)}{d\varphi}\,,
\end{align}
respectively, where $g$ is the determinant of the metric and we have denoted  ${\cal L}_F=\partial {\cal L} _{\rm NLED}(F)/\partial F$.

\subsection{Spherically symmetric solutions}

For the sake of our considerations we shall work in the static, spherically symmetric, line element 
\begin{equation}
ds^2=A(r)dt^2-B(r)dr^2-\Sigma^2(r) \, d\Omega^2,\label{m}
\end{equation}
where $A(r), B(r)$ and $\Sigma(r)$ are functions of the radial coordinate $r$ and $d\Omega^2\equiv d\theta^{2}+\sin^{2}\left(\theta\right)d\phi^{2}$
is the line element on the two-spheres. We point out that in this line element we have included a non-trivial radial function $\Sigma^2(r)$ given the fact that we are interested in geometries implementing a bouncing behaviour in the radial sector, for which $\Sigma^2(r)$ is non-monotonic. On the other hand, it is always possible to perform a change of variables of the form $B(r)dr^2=dx^2/A(r)$ to write the line element (\ref{m}) under a more canonical form with $g_{tt}=g^{xx}$, meaning that one of these two metric functions is redundant. Nonetheless, for the sake of generality, we shall work in the line element above and later fix one of the metric functions. 

In Ref. \cite{Harada:2021bte}, Harada found the following exact vacuum solution of the CG field equations  \eqref{eq_CG} as
\begin{equation}
    A(r)=\frac{1}{B(r)}=1 - \frac{2M}{r} - \frac{\Lambda}{3}r^2 + \gamma r. \label{vacuum_sol}
\end{equation}
where $M$ is the standard mass term of the Schwarzschild solution (to which the solution above reduces when $\Lambda=0$ and the parameter $\gamma$ resulting from the theory goes to zero, $\gamma \to 0$). Furthermore,  $\Lambda$ arises as an integration constant, and given the fact that the term $-\Lambda r^2/3$ denotes a de Sitter term makes $\Lambda$ naturally interpreted as a cosmological constant term. 

In the framework of CG, one may interpret that $\gamma$ acts as a crucial parameter that dictates the influence of the Cotton tensor on the gravitational field. Indeed, CG introduces the Cotton tensor as an additional component of the gravitational field, particularly affecting higher-order corrections to the curvature. Here, the parameter $\gamma$ dictates the relative strength of this correction and, therefore, its importance in describing deviations from GR’s predictions. For example, in the cosmological context, the parameter $\gamma$ can help address issues such as the accelerating expansion of the universe without invoking dark energy. Similarly, in galactic dynamics, $\gamma$ contributes to modifications that could explain phenomena such as the galaxy rotation curves without requiring dark matter, as discussed in 
\cite{Harada:2021bte,Harada:2022edl}.

In fact, the linear term in the radial coordinate, denoted the Rindler term \cite{Grumiller:2010bz}, leads to an anomalous acceleration in geodesics of test-particles. In \cite{Grumiller:2010bz}, the Rindler term appears as a linear potential that contributes to deviations from GR, and is included to explore the effects of additional forces or modifications to gravity at large scales. It was argued, as mentioned above, that such a term might be connected to phenomenological attempts to account for dark matter or dark energy without invoking new particles, by modifying the gravitational field itself \cite{Grumiller:2010bz}. In essence, this linear term represents a Rindler-like acceleration, which introduces a uniform acceleration in the system, analogous to the behavior in Rindler coordinates that describe a uniformly accelerated observer. This approach can offer insights into deviations from Newtonian gravity and GR, especially in the weak-field or large-scale limits.

Our aim is to generalize the above solution in CG to implement bouncing behaviours in the radial function. To this end we shall take the canonical choice, due to Ellis \cite{Ellis:1973yv}, as
\begin{equation} \label{eq:bb}
    \Sigma(r)=\sqrt{L_0^2+r^2}\,,
\end{equation}
where the parameter $L_0 \in \Re$ is a regularization parameter and possesses the dimension of a length. This constant can be consistently interpreted as a magnetic charge $q$, provided that we source the line element with a NLED and scalar fields. This means that the coordinate $r$ extends over the whole real line, $r \in (-\infty,+\infty)$, while the function $\Sigma(r)$ which gives the area of the two-spheres as $S=4\pi \Sigma^2(r)$ takes a minimum value $\Sigma(r)=q$ at $r=0$.

The invariants associated to such fields are
\begin{equation}
    F_{23}=q \sin\theta \,,
\end{equation}
for the electromagnetic field, and 
\begin{equation}
    F=\frac{q^2}{2 \Sigma^4(r)}.
\end{equation}
for the scalar one. In addition, to assess the consistency of our solutions we shall also use the following useful relationship
\begin{equation}
    {\cal L}_F=\frac{\partial {\cal L} _{\rm NLED}}{\partial r} \bigg(\frac{\partial F}{\partial r}\bigg)^{-1} \ .\label{RC}
\end{equation}

The field equations \eqref{eq_CG} are third-order coupled gravitational equations for the metric, which makes it very difficult to obtain a solution by direct integration when the matter content is included. To overcome this difficulty and find new solutions for these equations, we use the following approach in our work: we propose the spherical symmetry for the metric as well as a functional form that already contains the new characteristic term of the vacuum solution in CG while at the same time implementing the bouncing behaviour given by Eq. (\ref{eq:bb}), and then integrate the equations of motion to determine the Lagrangian of the NLED as well as its derivative in order to check the consistency equation \eqref{RC}. In this way, we will be able to generalize some known bouncing solutions found within GR (namely, the Simpson-Visser and Bardeen solutions), and obtain the specific shapes of the matter (electromagnetic and scalar),

To run this procedure we first solve the $(0,0,1)$ and $(2,1,2)$ components of the CG field equations \eqref{eq_CG}. As described below, respectively
\begin{widetext}
\begin{align}
&
-\frac{1}{6A(r)^{2}B(r)^{3}\left(q^{2}+r^{2}\right)^{3}}\Bigg\{A(r)B(r)\left(q^{2}+r^{2}\right)^{2}A'(r)\Big[A'(r)\left(2\left(q^{2}+r^{2}\right)B'(r)-rB(r)\right)-4B(r)\left(q^{2}+r^{2}\right)A''(r)\Big]
\nonumber
\\
& +2A(r)^{3}\Big\{
-2r\left(q^{2}+r^{2}\right)^{2}B'(r)^{2}+B(r)\left(q^{2}+r^{2}\right)\Big[r\left(q^{2}+r^{2}\right)B''(r)+3q^{2}B'(r)\Big]+2B(r)^{2}\left(3q^{2}r+r^{3}\right)
\nonumber
\\
&
-2rB(r)^{3}\left(q^{2}+r^{2}\right) \Big\} +2B(r)^{2}\left(q^{2}+r^{2}\right)^{3}A'(r)^{3}+
A(r)^{2}\left(q^{2}+r^{2}\right)\Big\{
2 \left(q^2+r^2\right)^2 A'(r) B'(r)^2+B(r)\left(q^{2}+r^{2}\right)\times
\nonumber
\\
&
\times\left[-3\left(q^{2}+r^{2}\right)A''(r)B'(r)-A'(r)\left(\left(q^{2}+r^{2}\right)B''(r)+rB'(r)\right)\right] +2B(r)^{2}\Big[\left(q^{2}+r^{2}\right)\left(A^{(3)}(r)\left(q^{2}+r^{2}\right)+2rA''(r)\right)
\nonumber
\\
&
-\left(q^{2}+2r^{2}\right)A'(r)\Big]\Big\}\Bigg\}
-\frac{\kappa^{2}}{6B(r)^{2}}\Bigg\{
-3\epsilon B(r)A'(r)\varphi'(r)^{2}+\epsilon A(r)B'(r)\varphi'(r)^{2} -2\epsilon A(r)B(r)\varphi'(r)\varphi''(r)
\nonumber
\\
&
+
\frac{A(r)B(r)^{2}\left[\left(q^{2}+r^{2}\right)\left(\left(q^{2}+r^{2}\right)^{2}L'(r)-2q^{2}{\cal L}_{F}'(r)+\left(q^{2}+r^{2}\right)^{2}V'(r)\right)+8q^{2}r{\cal L}_{F}(r)\right]}{\left(q^{2}+r^{2}\right)^{3}}\Bigg\}=0,
\end{align}

\begin{align}
&
\frac{1}{12 A(r)^3 B(r)^3 \left(q^2+r^2\right)^2}\Bigg\{A(r)B(r)\left(q^{2}+r^{2}\right)^{2}A'(r)\Big[A'(r)\left(2\left(q^{2}+r^{2}\right)B'(r)-rB(r)\right)-4B(r)\left(q^{2}+r^{2}\right)A''(r)\Big]
\nonumber
\\
& +2A(r)^{3}\Big\{
-2r\left(q^{2}+r^{2}\right)^{2}B'(r)^{2}+B(r)\left(q^{2}+r^{2}\right)\Big[r\left(q^{2}+r^{2}\right)B''(r)+3q^{2}B'(r)\Big]+2B(r)^{2}\left(3q^{2}r+r^{3}\right)
\nonumber
\\
&
-2rB(r)^{3}\left(q^{2}+r^{2}\right) \Big\} +2B(r)^{2}\left(q^{2}+r^{2}\right)^{3}A'(r)^{3}+
A(r)^{2}\left(q^{2}+r^{2}\right)\Big\{
2 \left(q^2+r^2\right)^2 A'(r) B'(r)^2+B(r)\left(q^{2}+r^{2}\right)\times
\nonumber
\\
&
\times\left[-3\left(q^{2}+r^{2}\right)A''(r)B'(r)-A'(r)\left(\left(q^{2}+r^{2}\right)B''(r)+rB'(r)\right)\right] +2B(r)^{2}\Big[\left(q^{2}+r^{2}\right)\left(A^{(3)}(r)\left(q^{2}+r^{2}\right)+2rA''(r)\right)
\nonumber
\\
&
-\left(q^{2}+2r^{2}\right)A'(r)\Big]\Big\}\Bigg\}
-\frac{1}{6}\kappa^{2}
\Bigg\{
\frac{\left(q^{2}+r^{2}\right)\left[\left(q^{2}+r^{2}\right)^{2}{\cal L}_{\text{NLED}}'(r)+q^{2}{\cal L}_{F}'(r)+\left(q^{2}+r^{2}\right)^{2}V'(r)\right]-q^{2}r{\cal L}_{F}(r)}{\left(q^{2}+r^{2}\right)^{2}}+
\nonumber
\\
&
\left.
+\frac{\epsilon\left(q^{2}+r^{2}\right)B'(r)\varphi'(r)^{2}}{B(r)^{2}}
-\frac{2\epsilon\varphi'(r)\left[\left(q^{2}+r^{2}\right)\varphi''(r)+3r\varphi'(r)\right]}{B(r)}\right\}=0,
\end{align}
\end{widetext}
where $A^{(3)}(r)$ denotes the third derivative with respect to $r$.
\color{black} This provides cumbersome yet analytic expressions for both the NLED Lagrangian density and its derivative as
\begin{widetext}
\begin{eqnarray}
&&{\cal L}_{\rm NLED}(r)= f_{0}+\frac{2f_{1}q^{2}}{\sqrt{q^{2}+r^{2}}}+\int\frac{1}{2\kappa^{2}q^{2}A(r)^{3}B(r)^{3}\left(q^{2}+r^{2}\right)^{5/2}}\Bigg\{2B(r)^{2}\left(q^{2}+r^{2}\right)^{3}A'(r)^{3}
\nonumber\\
&&
+\left(q^{2}+r^{2}\right)^{2}A(r)B(r)A'(r)\Big[A'(r)\left(2\left(q^{2}+r^{2}\right)B'(r)-rB(r)\right)-4B(r)\left(q^{2}+r^{2}\right)A''(r)\Big]
\nonumber\\
&&	
+2A(r)^{3}\left[-2rB(r)^{3}\left(q^{2}+r^{2}\right)-2r\left(q^{2}+r^{2}\right)^{2}B'(r)^{2}+2rB(r)^{2}\left(\kappa^{2}\epsilon\left(q^{2}+r^{2}\right)^{2}\varphi'(r)^{2}+3q^{2}+r^{2}\right)
	\right.
\nonumber\\
&&
+B(r)\left(q^{2}+r^{2}\right)\left(r\left(q^{2}+r^{2}\right)B''(r)+3q^{2}B'(r)\right)\Big]
-A(r)^{2}\left(q^{2}+r^{2}\right)\Bigg[-2\left(q^{2}+r^{2}\right)^{2}A'(r)B'(r)^{2}
\nonumber\\
&&	
+B(r)\left(q^{2}+r^{2}\right)\left(3\left(q^{2}+r^{2}\right)A''(r)B'(r)+A'(r)\left(\left(q^{2}+r^{2}\right)B''(r)+rB'(r)\right)\right)
\nonumber\\
&&
+2B(r)^{2}\left(A'(r)\left(\kappa^{2}\epsilon\left(q^{2}+r^{2}\right)^{2}\varphi'(r)^{2}+q^{2}+2r^{2}\right)-\left(q^{2}+r^{2}\right)\left(A^{(3)}(r)\left(q^{2}+r^{2}\right)+2rA''(r)\right)\right)\Bigg] \Bigg\} dr\nonumber \\
&&+\int\frac{1}{\kappa^{2}A(r)^{3}B(r)^{3}\left(q^{2}+r^{2}\right)^{3}}\bigg\{-2B(r)^{2}\left(q^{2}+r^{2}\right)^{3}A'(r)^{3}
	\nonumber\\
&&	-\left(q^{2}+r^{2}\right)^{2}A(r)B(r)A'(r)\left[A'(r)\left(2\left(q^{2}+r^{2}\right)B'(r)-rB(r)\right)-4B(r)\left(q^{2}+r^{2}\right)A''(r)\right]
	\nonumber\\
&&	-A(r)^{3}\left[-4r\left(q^{2}+r^{2}\right)^{2}B'(r)^{2}+B(r)^{3}\left(q^{2}+r^{2}\right)\left(\kappa^{2}\left(q^{2}+r^{2}\right)^{2}V'(r)-4r\right)
	\right.\nonumber\\
&&	+B(r)\left(q^{2}+r^{2}\right)\left(2r\left(q^{2}+r^{2}\right)B''(r)+B'(r)\left(\kappa^{2}\epsilon\left(q^{2}+r^{2}\right)^{2}\varphi'(r)^{2}+6q^{2}\right)\right)
	\nonumber\\
&&	\left.+2B(r)^{2}\left(2r\left(3q^{2}+r^{2}\right)-\kappa^{2}\epsilon\left(q^{2}+r^{2}\right)^{3}\varphi'(r)\varphi''(r)\right)\right]
	\nonumber\\
&&	+A(r)^{2}\Bigg[-2\left(q^{2}+r^{2}\right)^{2}A'(r)B'(r)^{2}-2B(r)^{2}\left(q^{2}+r^{2}\right)\left(A^{(3)}(r)\left(q^{2}+r^{2}\right)+2rA''(r)\right)\nonumber\\
&&	+B(r)\left(q^{2}+r^{2}\right)\left(3\left(q^{2}+r^{2}\right)A''(r)B'(r)+A'(r)\left(\left(q^{2}+r^{2}\right)B''(r)+rB'(r)\right)\right)
	\nonumber\\
&&	+B(r)^{2}A'(r)\left(3\kappa^{2}\epsilon\left(q^{2}+r^{2}\right)^{2}\varphi'(r)^{2}+2\left(q^{2}+2r^{2}\right)\right)]\left(q^{2}+r^{2}\right)\Bigg]\Bigg\} dr,	\label{L_BB} 
 \end{eqnarray}
and 
 \begin{eqnarray}
&& {\cal L}_F(r) = \left(q^2+r^2\right)^{3/2}\Bigg\{f_0+\int\frac{1}{2\kappa^{2}q^{2}A(r)^{3}B(r)^{3}\left(q^{2}+r^{2}\right)^{5/2}}\Bigg\{2B(r)^{2}\left(q^{2}+r^{2}\right)^{3}A'(r)^{3}
\nonumber\\
&&
+\left(q^{2}+r^{2}\right)^{2}A(r)B(r)A'(r)\Big[A'(r)\left(2\left(q^{2}+r^{2}\right)B'(r)-rB(r)\right)-4B(r)\left(q^{2}+r^{2}\right)A''(r)\Big]
\nonumber\\
&&	
+2A(r)^{3}\left[-2rB(r)^{3}\left(q^{2}+r^{2}\right)-2r\left(q^{2}+r^{2}\right)^{2}B'(r)^{2}+2rB(r)^{2}\left(\kappa^{2}\epsilon\left(q^{2}+r^{2}\right)^{2}\varphi'(r)^{2}+3q^{2}+r^{2}\right)
	\right.
\nonumber\\
&&
+B(r)\left(q^{2}+r^{2}\right)\left(r\left(q^{2}+r^{2}\right)B''(r)+3q^{2}B'(r)\right)\Big]
-A(r)^{2}\left(q^{2}+r^{2}\right)\Bigg[-2\left(q^{2}+r^{2}\right)^{2}A'(r)B'(r)^{2}
\nonumber\\
&&	
+B(r)\left(q^{2}+r^{2}\right)\left(3\left(q^{2}+r^{2}\right)A''(r)B'(r)+A'(r)\left(\left(q^{2}+r^{2}\right)B''(r)+rB'(r)\right)\right)
\nonumber\\
&&
+2B(r)^{2}\left(A'(r)\left(\kappa^{2}\epsilon\left(q^{2}+r^{2}\right)^{2}\varphi'(r)^{2}+q^{2}+2r^{2}\right)-\left(q^{2}+r^{2}\right)\left(A^{(3)}(r)\left(q^{2}+r^{2}\right)+2rA''(r)\right)\right)\Bigg] \Bigg\} dr\Bigg\}.\label{LF_BB}
\end{eqnarray}
\end{widetext}
respectively, where $f_0$ and $f_1$ are integration constants.

As for the scalar field $\varphi$, we choose the corresponding expression of the GR case which is given by \cite{Rodrigues2023}
\begin{equation}
    \varphi(r)=\frac{\tan ^{-1}\left(r/q\right)}{\sqrt{\kappa ^2 (-\epsilon )}}.\label{phi_BB}
\end{equation}
Since the transformation of the radial coordinate $r$ in spacetime can be chosen freely, with the only restriction of being real and continuous, we will fix the gauge $B(r) = A(r)^{-1}$ to simplify all our expressions, including equations~\eqref{L_BB} and \eqref{LF_BB}. From this point onward, we will use this gauge, which is consistent with the area radius function in the angular part of the metric, as given in equation \eqref{eq:bb}.

After we substitute Eq. \eqref{phi_BB} into Eqs.~\eqref{L_BB} and~\eqref{LF_BB} and then into the condition \eqref{RC} with the choice (\ref{eq:bb}) we arrive at the equation
\begin{equation}
    \frac{\left(q^{2}+r^{2}\right)}{2r}\left[-\frac{2A'(r)}{\kappa^{2}}-\frac{\left(q^{2}+r^{2}\right)^{2}V'(r)}{q^{2}}\right]=0\,,
    \label{rc_BB}
\end{equation}
which yields the following solution for the potential:
\begin{equation}
    V(r)= V_0 + \int -\frac{2  q^2 A'(r)}{\kappa ^2\left(q^2+r^2\right)^2 }dr.\label{V_BB}
\end{equation}

This completes our framework to finding new solutions in CG gravity. In order to achieve it so we just need to set a functional form for the metric coefficient $A(r)$ (since as mentioned above we can always choose $B=A(r)^{-1}$ without any loss of generality). In this sense, in the following sections  \ref{BB_SV} and \ref{BB_Bardeen} we present the bouncing solutions for the generalizations of Simpson-Visser and Bardeen solutions. In this sense, we use the set of equations~\eqref{L_BB} and \eqref{LF_BB}  for the NLED Lagrangian density and its derivative, and \eqref{phi_BB} and \eqref{V_BB}  for the scalar field potential, in order to reconstruct the functional form of the matter field Lagrangians.

\section{Simpson-Visser type solution}\label{BB_SV}

\subsection{Metric and solutions}

A natural generalization of the solution (\ref{vacuum_sol}) is to imbue it with the choice (\ref{eq:bb}) for the radial function, that is, to propose the metric coefficient
\begin{equation}
    A(r)= 1-\frac{2M}{\left(q^{2}+r^{2}\right)^{1/2}}-\frac{1}{3}\Lambda\left(q^{2}+r^{2}\right)+\gamma\sqrt{q^{2}+r^{2}}.\label{a_BB}
\end{equation}
This is the counterpart of the Simpson-Visser black bounce introduced in \cite{Simpson:2018tsi}, which is recovered when the CG parameter $\gamma=0$ (and for vanishing cosmological parameter). The black bounce proposal via the choice (\ref{eq:bb}) has become canonical in  the literature given its ability to regularize curvature singularities in any spherically symmetric space of interest, like ours. Furthermore, it has been shown that it is generically supported by combinations of NLED and scalar fields, see e.g. \cite{Bronnikov:2021uta}. A similar approach is possible in our case, as shall be shown later in this section.

In order to analyze the properties of the metric function (\ref{a_BB}) we shall characterize the presence of horizons via the condition $g_{rr}^{-1}(r_H)=0$ (which coincides with the Killing horizon $g_{tt}=0$), that is:
\begin{equation}
     A(r_{H})=0,\label{rH}
\end{equation}
where the radius $r_{H}$ denotes the presence of the horizon. In order to analyze the number and nature of such horizons, we add the condition 
\begin{equation}
    \frac{d A(r_{H})}{dr}\bigg|_{r=r_H}=0,
    \label{der_a}
\end{equation}
which is capable to detect the extreme black holes (characterized by a degenerate horizon). Such a relation will numerically provide the critical values of the parameters of the model, namely, $M$, $q$, $\Lambda$ or $\gamma$ for which such extreme black holes exist. Elaborating from them we shall be able to characterize the remaining solutions within each choice of the parameters.  Given  the fact that there are four free parameters, in our approach we set three of them and take the remaining one (either $M$, $q$, or $\gamma$)  as a free parameter, and investigate the corresponding solutions for both positive and negative values of the cosmological constant $\Lambda$. This leads to six classes of solutions in order to explore all the possible geometrical structures within this setting.  For each of them we numerically determine the value of the critical values of the free parameter (either $M$, $q$, or $\gamma$) from the simultaneous resolution of Eqs. \eqref{rH} and~\eqref{der_a}.\\


{\it (i) Type-I solution: $\{q = 0.5, \Lambda = -0.2, \gamma=0.01\}$}

\begin{figure*}[htb!]
\centering
\subfigure[~Type-I solution: $\{q = 0.5, \Lambda = -0.2, \gamma=0.01\}$] 
{\label{figM1_BB1}\includegraphics[width=7.75cm]{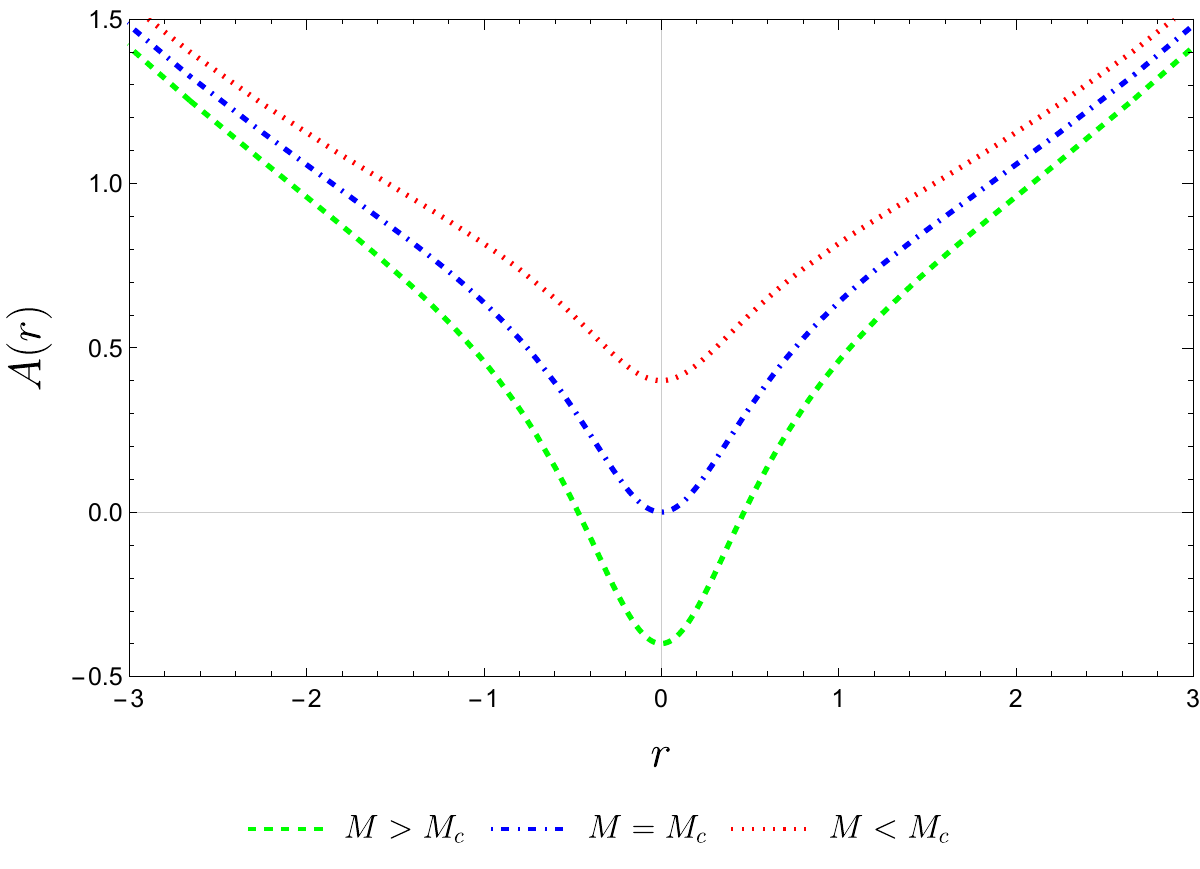} }
\hspace{0.75cm}
\subfigure[~Type-II solution: $\{q = 0.5, \Lambda = 0.2, \gamma=0.01\}$] 
{\label{figM1_BB2}\includegraphics[width=7.75cm]{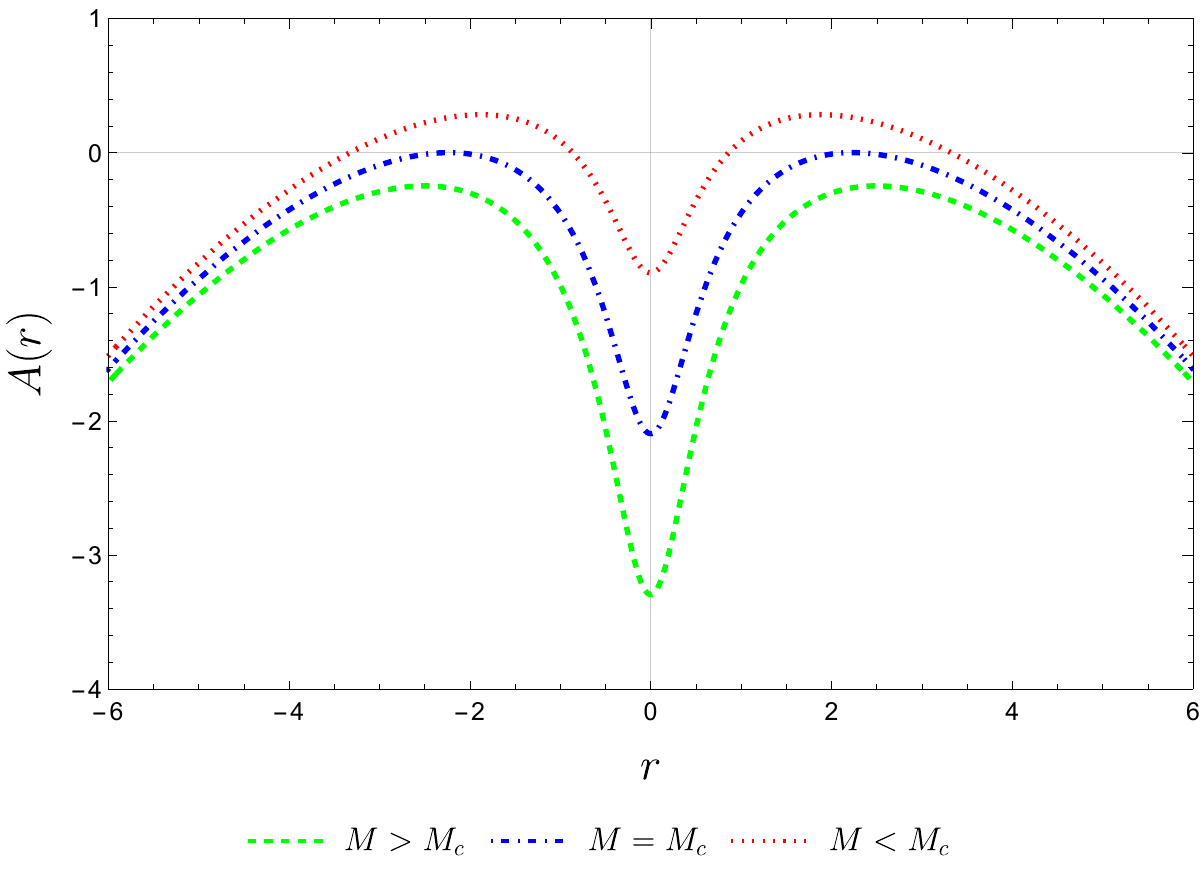} }
\subfigure[~Type-III solution: $\{\gamma=0.01,M=2.0,\Lambda=-0.2\}$] 
{\label{figM1_BB3}\includegraphics[width=7.75cm]{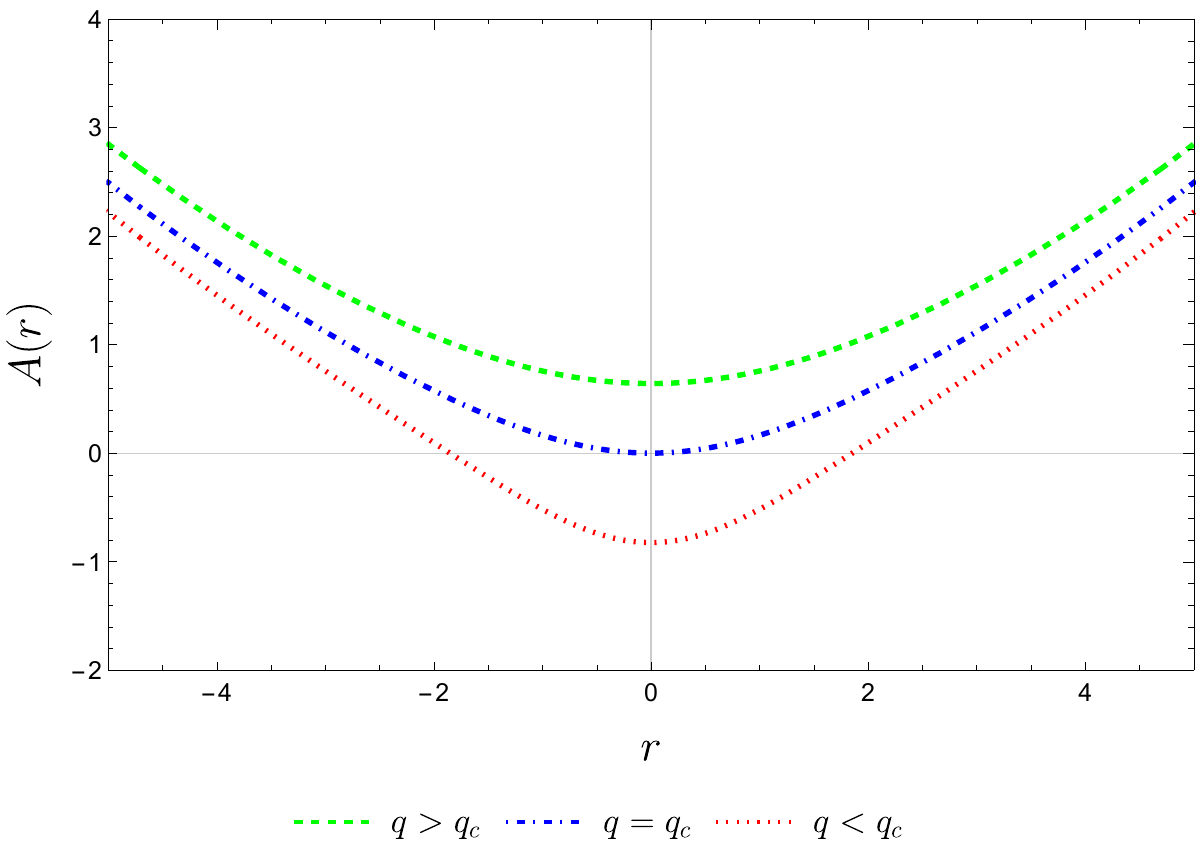}}
\hspace{0.75cm}
\subfigure[~Type-IV solution: $\{\gamma=0.5,M=2.0,\Lambda=0.2\}$] 
{\label{figM1_BB4}\includegraphics[width=7.75cm]{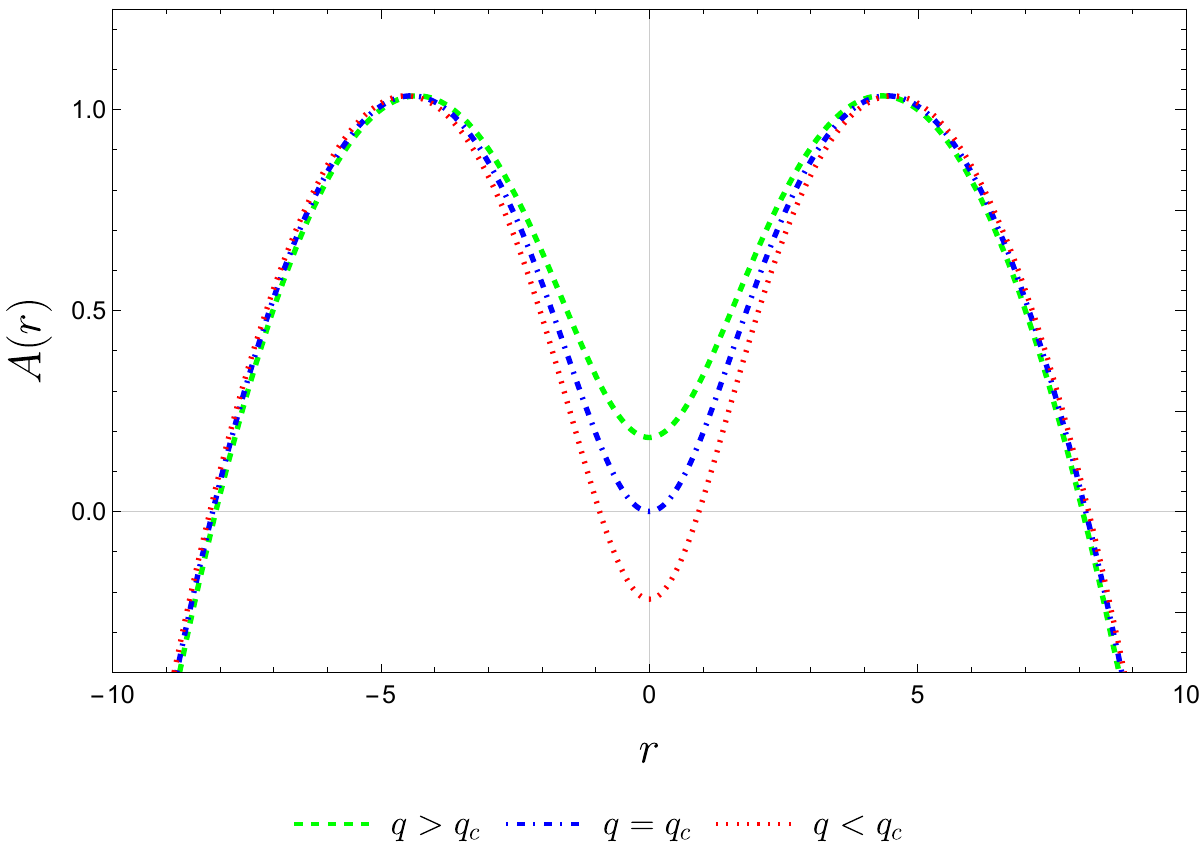}}
\subfigure[~Type-V solution: $\{q=0.75,M=1.0,\Lambda=-0.2\}$] 
{\label{figM1_BB5}\includegraphics[width=7.75cm]{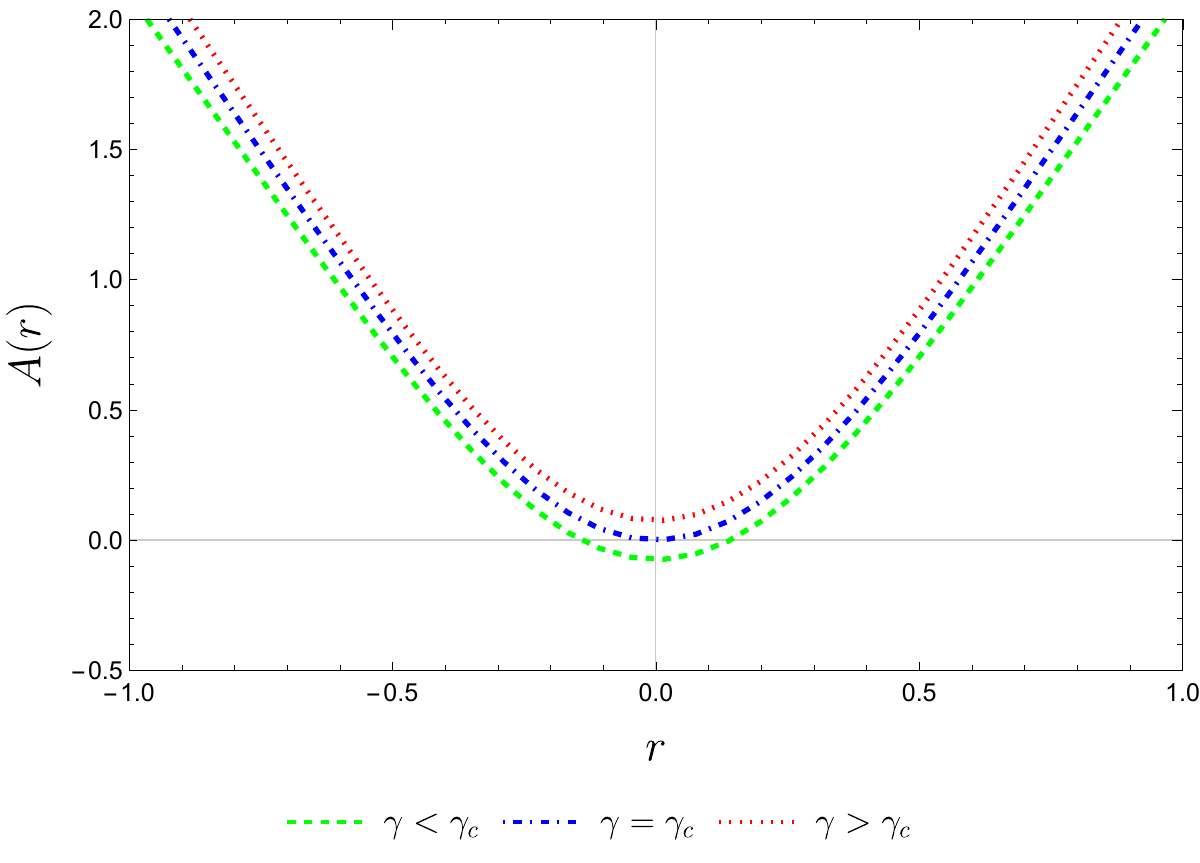}}
\hspace{0.75cm}
\subfigure[~Type-VI solution: $\{q=0.52,M=1.2,\Lambda=0.2\}$] 
{\label{figM1_BB6}\includegraphics[width=7.75cm]{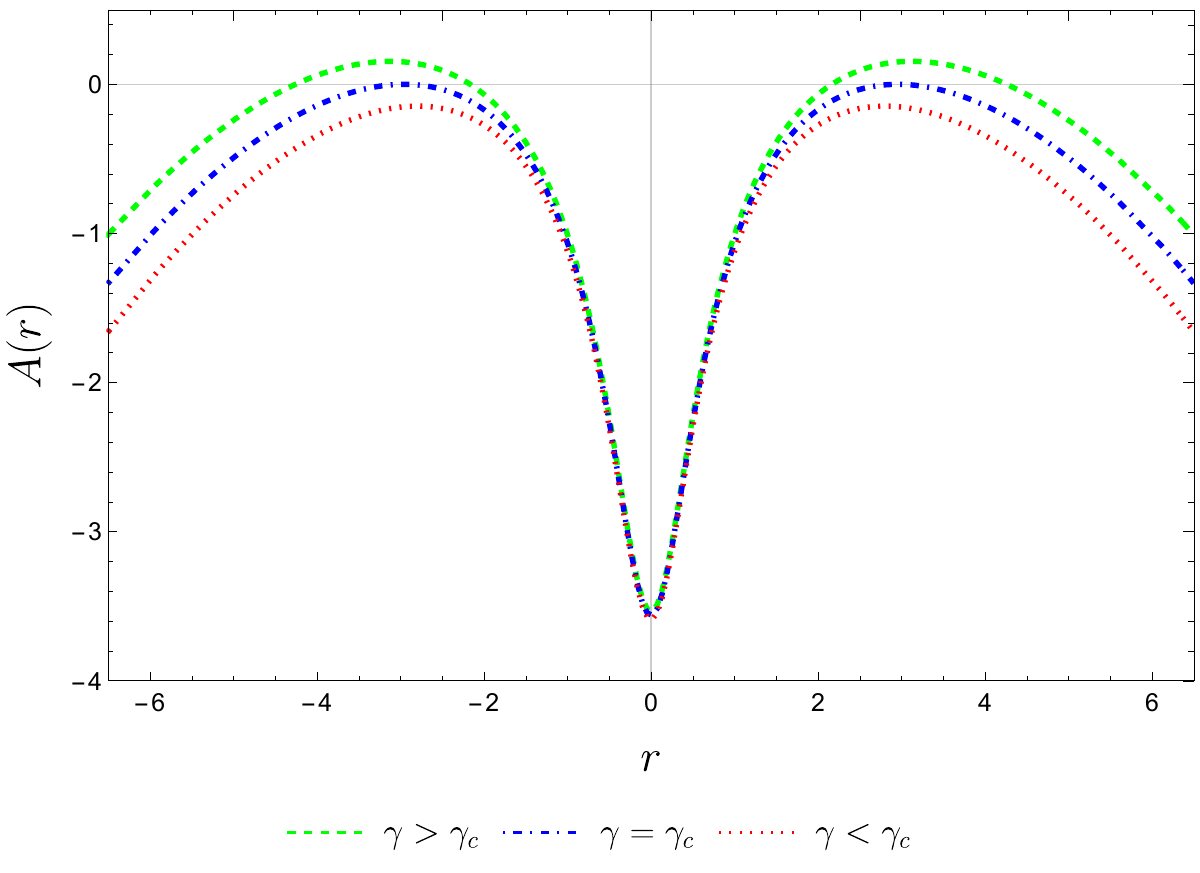}}
\caption{The plot depicts $A(r)$ for the six types of solutions described in Sec. \ref{BB_SV} (see definitions in the main text), found by combining values for the four parameters $\{M,q,\Lambda,\gamma\}$ characterizing the solutions. See the text for more details.}
\label{Fig3}
\end{figure*}

For these values the critical mass, found by solving simultaneously Eqs.(\ref{rH}) and (\ref{der_a}),  takes the value $M_c = 0.255$. Fig.~\ref{figM1_BB1} shows the behavior of the metric function~\eqref{a_BB} in terms of the radial coordinate $r$, for three mass scenarios: $M > M_c$, $M = M_c$, and $M < M_c$. When the mass exceeds the critical value, $M > M_c$, we observe the presence of two event horizons. This is possible given the fact that at $r=0$ we have a bounce from the region $r > 0$ to the region $r < 0$, and conversely, due to the symmetry with respect to $r$, a bounce from the region $r < 0$ to the region $r > 0$ is also possible. In the case where the mass is equal to the critical mass, $M = M_c$, we observe a degenerate double horizon at $r=0$, and a bounce is still present at $r=0$. Finally, when the mass is less than the critical mass, $M < M_c$, there is no horizon formation, and the geometry we have is a wormhole with a traversable throat at $r=0$. \\


{\it (ii) Type-II solution: $\{q = 0.5, \Lambda = 0.2, \gamma=0.01\}$}

In this case the critical mass is given by $M_c=0.770$. In the case where $M < M_c$, we see in Fig.~\ref{figM1_BB2} the formation of four horizons throughout the entire region $r\in(-\infty,+\infty)$, two on each side of the bounce. The innermost horizons on each side correspond to the event horizons, and the outermost are the cosmological horizons, denoted here by $r_{\Lambda}$. This is due to the fact that outside the cosmological horizons the metric function $A(r)$ is negative with signature for the metric ($-,+,-,-$). The bounce thus connects two black hole space-times both surrounded by a cosmological horizon. In the case where the mass is equal to the critical mass, $M = M_c$, we observe the existence of degenerate horizons on each side of the bounce, made up of the event and cosmological horizons. When the mass exceeds the critical value, $M > M_c$, there is no horizon of any kind, and the metric function remains negative throughout the entire $r$ region, which can thus be interpreted as a traversable wormhole embedded in a de Sitter space-time.\\


{\it (iii) Type-III solution: $\{\gamma=0.01,M=2.0,\Lambda=-0.2\}$}

Now the combined resolution of Eqs.(\ref{rH}) and (\ref{der_a}) provides instead a critical value for the charge as  $q_c=2.665$. The behavior of $A(r)$ is shown in Fig.~\ref{figM1_BB3} for the scenarios of $q > q_c$, $q=q_c$ and $q<q_c$.
Now, when $q>q_c$, there is no event horizon, and the space-time is described by a traversable wormhole with its throat at $r=0$. In the configuration where $q=q_c$, we have a case similar to Fig. \ref{figM1_BB1} for $M=M_c$ with a degenerate horizon at $r=0$. And when $q<q_c$, we find a situation similar to the dashed green curve represented in Fig.~\ref{figM1_BB1}, which corresponds to the scenario in which $M > M_c$ for which there are event horizons on both sides of $r=0$. \\


{\it (iv) Type-IV solution: $\{\gamma=0.5,M=2.0,\Lambda=0.2\}$}

The critical charge for these conditions reads $q_c=2.240$. The behavior of $A(r)$ as a function of $r$ is shown in Fig.~\ref{figM1_BB4}, where the charge takes  the values $q > q_c$, $q=q_c$, and $q < q_c$. When $q>q_c$, we observe the presence of two event horizons and two cosmological horizons; for $q=q_c$ event horizons and cosmological horizons merge to form a single degenerate horizon; while for $q<q_c$ then we have a traversable wormhole. This scenario is thus analogous to the case represented in Fig.~\ref{figM1_BB2} with $M<M_c$. \\


{\it (v) Type-V solution: $\{q=0.75,M=1.0,\Lambda=-0.2\}$}

In this case the joined resolution of Eqs.(\ref{rH}) and (\ref{der_a}) provides a critical value for the constant $\gamma_c=2.172$. The behavior of $A(r)$ is shown in  Fig.~\ref{figM1_BB5} for the scenarios $\gamma < \gamma_c$, $\gamma=\gamma_c$ and $\gamma>\gamma_c$. Now, when $\gamma>\gamma_c$, there is no event horizon and a traversable wormhole arises, similarly as the cases of $M<M_c$ and $q>q_c$ described in Figs. \ref{figM1_BB1} and \ref{figM1_BB3}, respectively.  
In the configuration where $\gamma=\gamma_c$, we have a case similar to Fig. \ref{figM1_BB1} for $M=M_c$ and also to the case of $q=q_c$ described in  Fig.~\ref{figM1_BB3}  (middle left). And when $\gamma<\gamma_c$, we find a situation similar to the curve represented in Fig.~\ref{figM1_BB1}, which corresponds to the scenarios of $M > M_c$ and also to the case of $q<q_c$ described in  Fig.~\ref{figM1_BB3}.\\


{\it (vi) Type-VI solution: $\{q=0.52,M=1.2,\Lambda=0.2\}$}

Under these conditions, we obtain $\gamma_c=0.133$. The behavior of $A(r)$ as a function of $r$ is shown in  Fig.~\ref{figM1_BB6},  where the parameter takes the values $\gamma> \gamma_c$, $\gamma=\gamma_c$ and $\gamma < \gamma_c$. In the case that $\gamma>\gamma_c$, we observe a similar behaviour as in  Fig.~\ref{figM1_BB2},  for $M<M_c$ and also for the case that $q < q_c$, as described in  Fig.~\ref{figM1_BB4}. In the case that $\gamma=\gamma_c$, we have a similar scenario as in  Fig.~\ref{figM1_BB2},  for $M= M_c$. Finally, for $\gamma < \gamma_c$, we have a wormhole geometry, analogous to the case in  Fig.~\ref{figM1_BB2}, but with $M>M_c$.

\subsection{Kretschmann scalar}

On the other hand we can study the regularity properties of our model by computing the  Kretschmannn scalar as
\begin{widetext}
\begin{align}
    &K=	
\frac{1}{3\left(q^{2}+r^{2}\right)^{5}}\Big\{36 M^2 \left(3 q^4-4 q^2 r^2+4 r^4\right)+	12 M q^2\Big[q^2 \left(r^2 \left(4 \Lambda  \sqrt{q^2+r^2}-9 \gamma \right)-8 \sqrt{q^2+r^2}\right)
	\nonumber\\
&+8 r^2 \sqrt{q^2+r^2}-2 r^4 \left(\gamma -\Lambda  \sqrt{q^2+r^2}\right)+q^4 \left(2 \Lambda  \sqrt{q^2+r^2}-7 \gamma \right)\Big]
	\nonumber\\
&+\left(q^{2}+r^{2}\right)^{}\left[4\Lambda^{2}q^{8}+8r^{6}\left(3\gamma^{2}-3\gamma\Lambda\sqrt{q^{2}+r^{2}}+\Lambda^{2}r^{2}\right)+q^{6}\left(27\gamma^{2}-20\gamma\Lambda\sqrt{q^{2}+r^{2}}+16\Lambda\left(\Lambda r^{2}-1\right)\right)
	\right. \nonumber
	\\
&\left.\left.4q^{2}r^{4}\left(12\gamma^{2}-13\gamma\Lambda\sqrt{q^{2}+r^{2}}+2\Lambda\left(3\Lambda r^{2}-1\right)\right)+q^{4}\left(-48\gamma\sqrt{q^{2}+r^{2}}\left(\Lambda r^{2}-1\right)+51\gamma^{2}r^{2}+4\left(7\Lambda^{2}r^{4}-6\Lambda r^{2}+9\right)\right)\right]\right\} 	.
	\label{K_BB}
\end{align}
\end{widetext}
which, at asymptotic infinity takes the value $R(r \to \infty) \to \frac{8 \Lambda ^2}{3}$, while being regular at the bouncing region $r=0$, as is transparent from Eq. \eqref{K_BB}.

\subsection{Matter Lagrangian reconstruction}

The above solutions can be shown to be supported by a NLED plus a scalar field matter field. To obtain the explicit shapes of their Lagrangian densities, we use Eq. \eqref{a_BB}  to obtain $\cal{L}_{\rm NLED}$ as
\begin{align}
{\cal L}_{\rm NLED}(r) &=  f_{0}+\frac{2f_{1}q^{2}}{\sqrt{q^{2}+r^{2}}}-\frac{2q^{2}}{6\kappa^{2}\sqrt{q^{2}+r^{2}}\left(q^{2}+r^{2}\right)^{2}}\times
		\nonumber \\
& \hspace{-0.5cm} \times\left[-45M-3\gamma\left(q^{2}+r^{2}\right)+4\sqrt{q^{2}+r^{2}}\right]+
\nonumber\\
&\hspace{-1.4cm}\frac{2q^{2}}{\kappa^{2}}\left(-\frac{6M}{\left(q^{2}+r^{2}\right)^{5/2}}-\frac{\gamma}{3\left(q^{2}+r^{2}\right)^{3/2}}+\frac{1}{2\left(q^{2}+r^{2}\right)^{2}}\right),
\label{L2_BB} 
\end{align}
and its derivative ${\cal L}_F$ as
\begin{eqnarray}
{\cal L}_F(r) &=&\left(q^{2}+r^{2}\right)^{3/2}\left[f_{1}-\left(-45M-3\gamma\left(q^{2}+r^{2}\right)
	\right.\right.\nonumber\\
&&\left.\left. +4\sqrt{q^{2}+r^{2}}\right)\Big/6\kappa^{2}\left(q^{2}+r^{2}\right)^{2}\right],\label{LF2_BB}
\end{eqnarray}
while for the scalar field  $\varphi(r)$ we use Eq.  (\ref{phi_BB}) to find the potential $V(r)$ as
\begin{equation}
V(r)= \frac{2 q^2 \left[6 M+5 \left(q^2+r^2\right) \left(\gamma -\Lambda  \sqrt{q^2+r^2}\right)\right]}{15 \kappa ^2 \left(q^2+r^2\right)^{5/2}}\,.
\label{V2_BB}
\end{equation}

On the other hand, inverting numerically the expressions above to obtain  $r(F)$ and $r(\cal{\varphi})$, we determine
\begin{eqnarray}
   {\cal L} _{\rm NLED} (F)  &=&	f_{0}+2\sqrt[4]{2}f_{1}\sqrt[4]{F}q^{3/2}-\frac{2F}{3\kappa^{2}}
	\nonumber \\   
 && +\frac{6\sqrt[4]{2}F^{5/4}M}{\kappa^{2}\sqrt{q}}+\frac{2^{3/4}\gamma F^{3/4}\sqrt{q}}{3\kappa^{2}},
 \label{L3_BB} 
\end{eqnarray}   
and
\begin{eqnarray}
V(\varphi) &=& \frac{2\left(3M\cos\left(2\text{\ensuremath{\varphi}}\sqrt{\kappa^{2}(-\epsilon)}\right)+3M\right)}{15\kappa^{2}\left(q^{2}\sec^{2}\left(\text{\ensuremath{\varphi}}\sqrt{\kappa^{2}(-\epsilon)}\right)\right)^{3/2}}\nonumber\\
&&+\frac{2q^{2}\left(\gamma-\Lambda\sqrt{q^{2}\sec^{2}\left(\text{\ensuremath{\varphi}}\sqrt{\kappa^{2}(-\epsilon)}\right)}\right)}{3\kappa^{2}\left(q^{2}\sec^{2}\left(\text{\ensuremath{\varphi}}\sqrt{\kappa^{2}(-\epsilon)}\right)\right)^{3/2}}.\label{V3_BB}
\end{eqnarray}  
Considering $f_0=0$, $f_1=0$, $\gamma=0$ and $\Lambda=0$ in the Lagrangian \eqref{L2_BB} we obtain the same expression as the one of GR as found in Ref. \cite{Rodrigues2023}. On the other hand, in expression \eqref{V3_BB} the scalar field must necessarily be a phantom field, i.e. 
$\epsilon=-1$, as is transparent in Eq. (\ref{phi_BB}). Note that in the limit $F\gg 1$, the NLED Lagrangian behaves as
\begin{align}
    {\cal L} _{\rm NLED} (F)\sim& f_{0}+2\sqrt[4]{2}f_{1}q^{3/2}F^{1/4}+\frac{6\sqrt[4]{2}F^{5/4}M}{\kappa^{2}\sqrt{q}}
    	\nonumber\\
   &+\frac{2^{3/4}\gamma q^{1/2}F^{3/4}}{3\kappa^{2}}-\frac{2F}{3\kappa^{2}}.
\end{align}
which is a non-Maxwellian behavior in the central region of the field. The bottom line of this analysis is that, similarly as in the case of GR \cite{Rodrigues2023}, the Simpson-Visser solution can be modelled as a solution of a NLED and a scalar field, though with a more general functional form.

\begin{figure}[t!]
\includegraphics[scale=0.48]{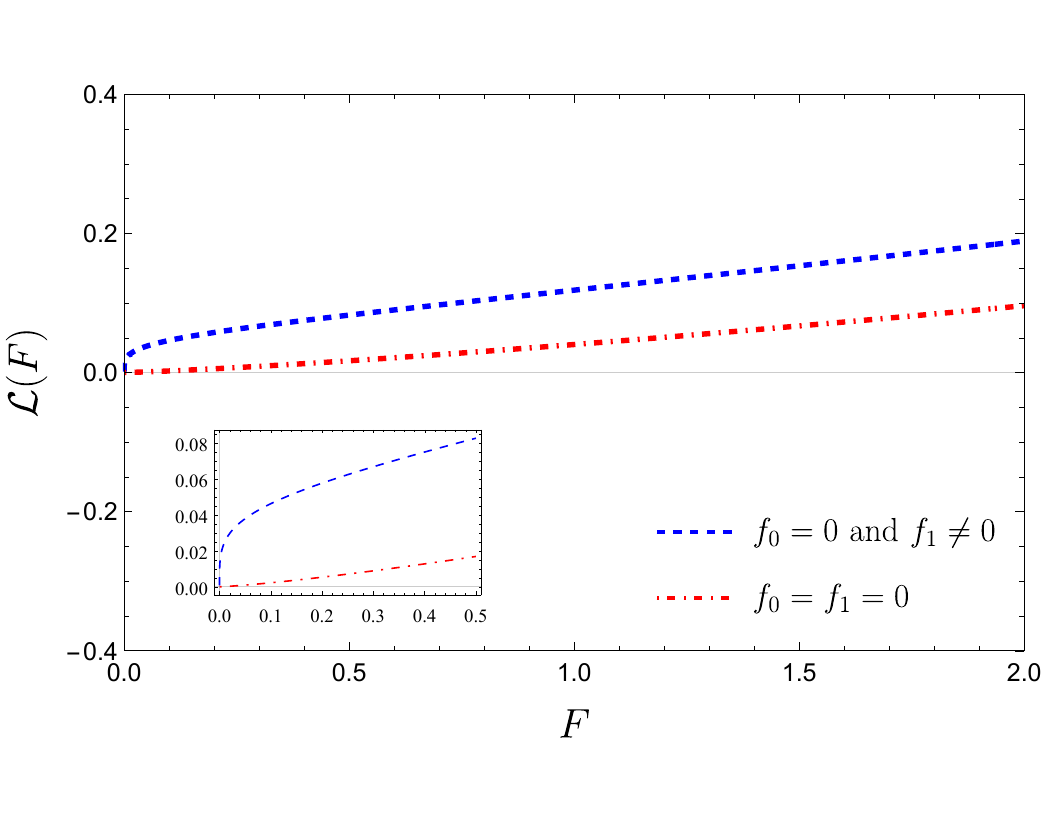}
\caption{The NLED Lagrangian ${\cal L}(F) $, given by Eq.~\eqref{L3_BB}. The blue dashed curve represents the behavior of ${\cal L}(F)$ versus $F$ with $f_0=0$ and $f_1=0.2$, while the red dotted-dashed curve corresponds to $f_0=f_1=0$. We have taken the values of the constants as  $\{\lambda=0,\Lambda=-0.2,q=0.3,M=2.0\}$. } 
\label{LxF_BB}
\end{figure}

\begin{figure}[t!]
\includegraphics[scale=0.55]{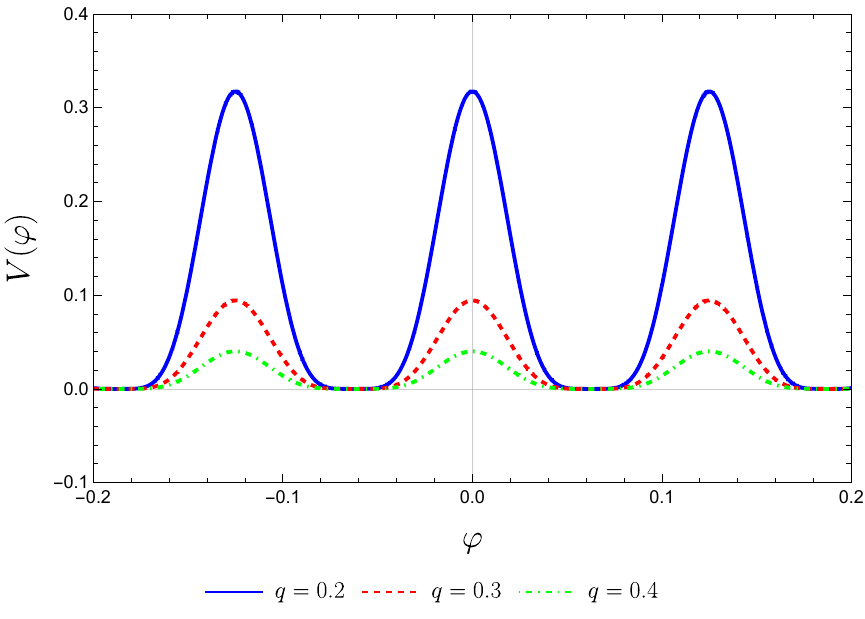}
\caption{The scalar field potential $V(\varphi)$, described by Eq.~\eqref{V3_BB} for different values of the charge taking $\epsilon=-1$,  for the values $\{\lambda=0,\Lambda=-0.2,M=2.0\}$.} 
\label{Vxphi_BB}
\end{figure}

In Fig. \ref{LxF_BB} we depict the behaviour of the NLED Lagrangian density given by Eq.~\eqref{L3_BB} using the blue dashed curve for $f_0=0$ and $f_1=0.2$ and for the case similar to GR with the red dotted-dashed curve with the values $f_0=0$ and $f_1=0$. Similarly, in Fig.~\ref{Vxphi_BB}. we illustrate the behaviour of the scalar field potential in Eq.~\eqref{V3_BB} for some choices of the model parameters. Note that the amplitude of this potential changes depending on the charge values. We illustrate this for three different values of the charge: $q = 0.2$, $q = 0.3$ and $q = 0.4$.

\section{Bardeen-type solution}\label{BB_Bardeen}

\subsection{Metric and solutions}

Consider now the following metric function:
\begin{equation}
    A(r)= 1-\frac{2Mr^{2}}{\left(q^{2}+r^{2}\right)^{3/2}}-\frac{1}{3}\Lambda\left(q^{2}+r^{2}\right)+\gamma\sqrt{q^{2}+r^{2}}.\label{a2_BB}
\end{equation}
This is a generalization, within CG, of the so-called Bardeen magnetic monopole \cite{Bardeen,Ayon-Beato:2000mjt} via the CG parameter $\gamma$, using a similar trail of thought as in the Simpson-Visser solution of the previous section with the replacement of the radial function with the choice (\ref{eq:bb}). Bardeen's monopole avoids the presence of a singularity via its replacement by a de Sitter core, a canonical procedure within GR to achieve this end \cite{Ansoldi:2008jw}, here generalized to display a bouncing behaviour on its radial sector. In this section we shall study the properties of this metric function using again to this end Eqs.~\eqref{rH} and~\eqref{der_a} for the radius of the horizon and the critical parameters of the solution, and split the discussion also into six sub-cases.\\


\begin{figure*}[htb!]
\centering
\subfigure[~Type-I solution: $\{q = 0.5, \Lambda = -0.2,\gamma=0.01\}$] 
{\label{figM_Bardeen1}\includegraphics[width=7.75cm]{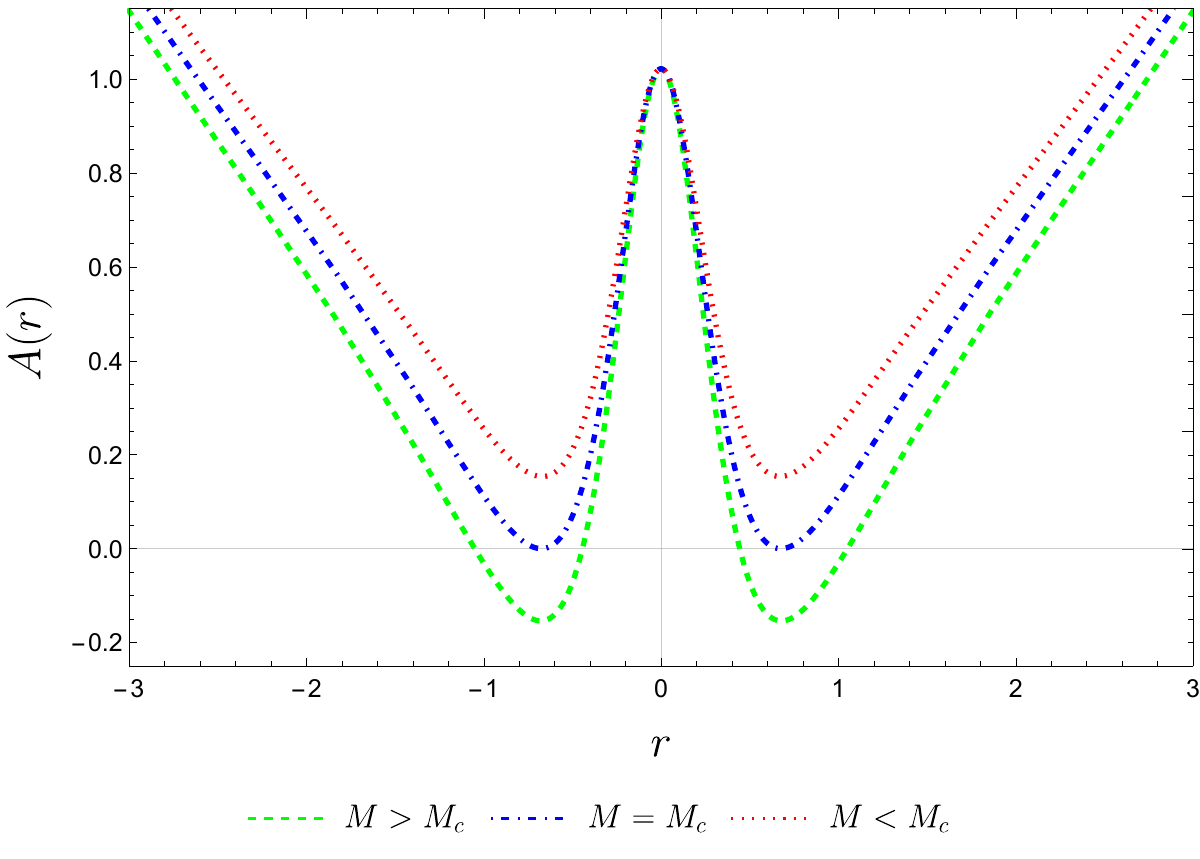} }
\hspace{0.75cm}
\subfigure[~Type-II solution: $\{q=0.5,\Lambda=0.2,\gamma=0.01\}$] 
{\label{figM_Bardeen2}\includegraphics[width=7.75cm]{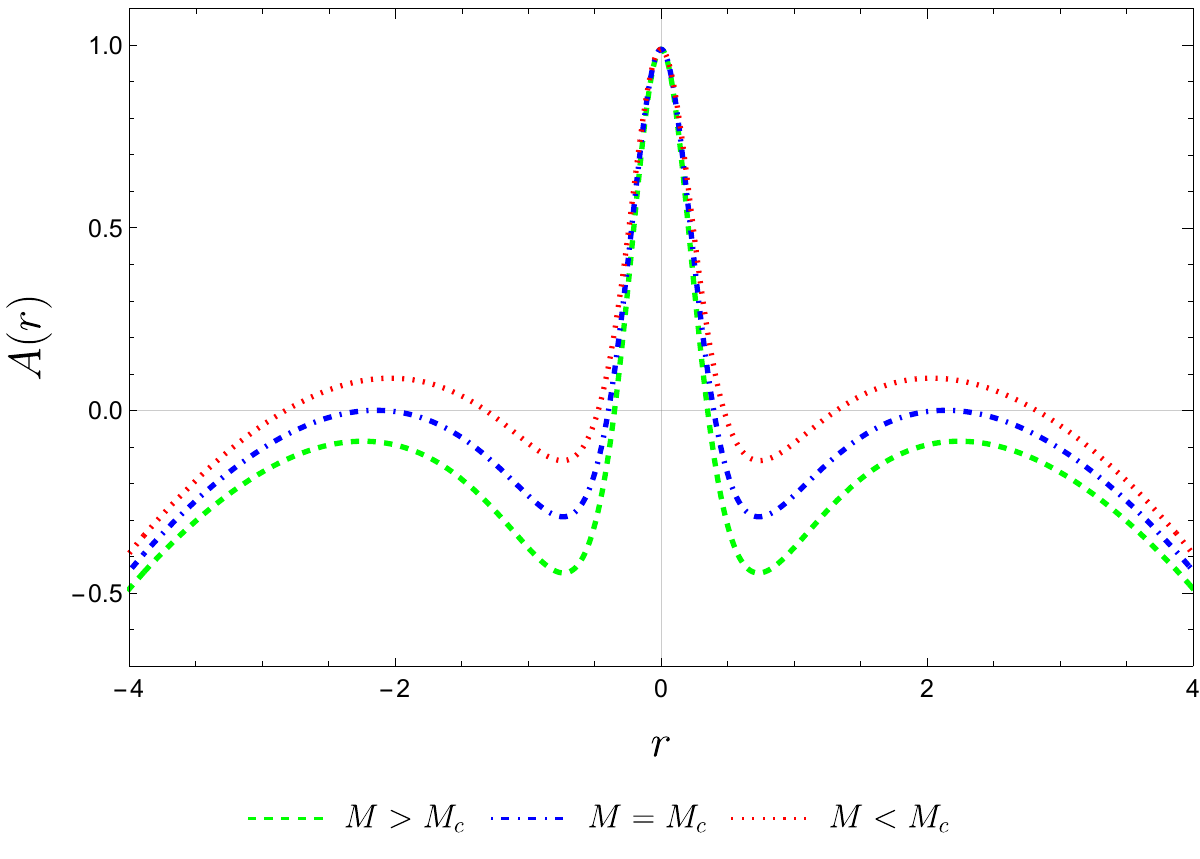} }
\subfigure[~Type-III solution: $\{\gamma=0.01,M=2.0,\Lambda=-0.2 \}$] 
{\label{figM_Bardeen3}\includegraphics[width=7.75cm]{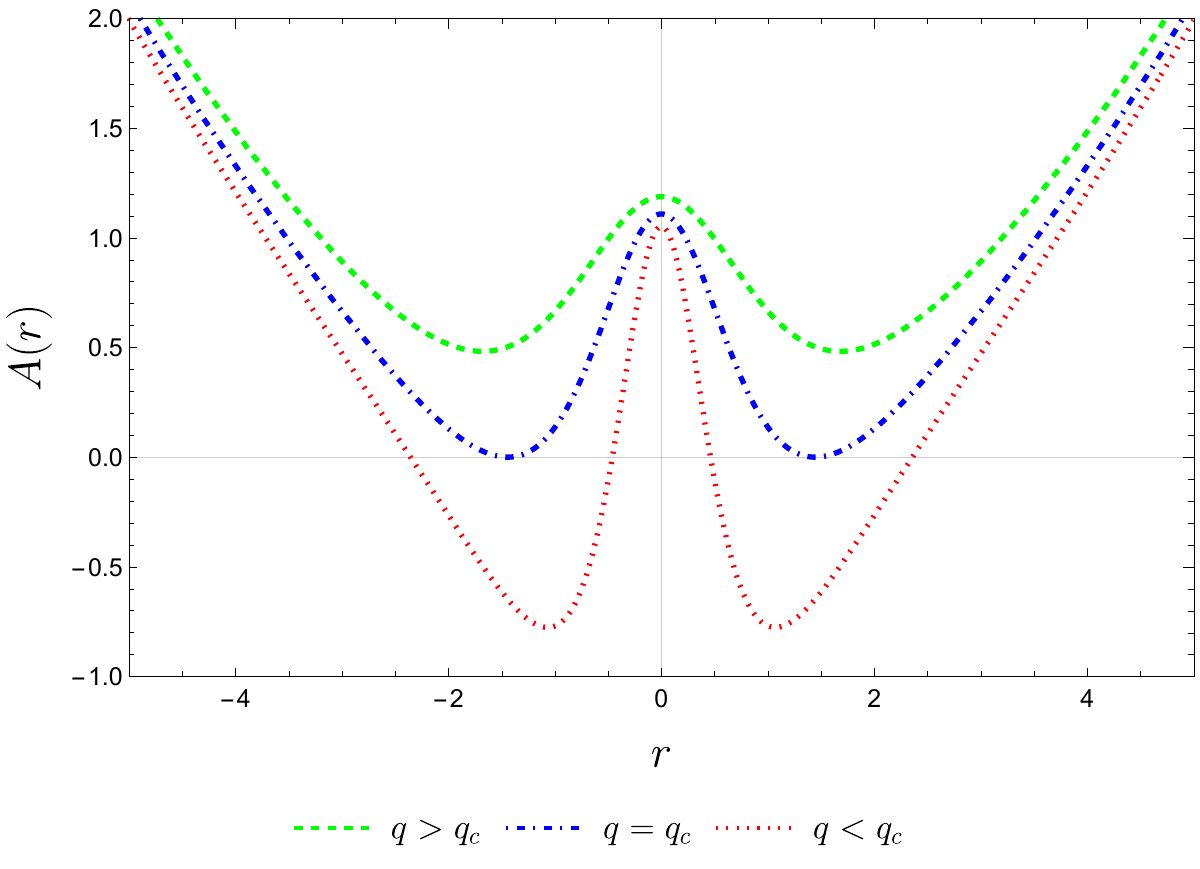}}
\hspace{0.75cm}
\subfigure[~Type-IV solution: $\{\gamma = 0.01,M = 0.5,\Lambda = 0.12\}$] 
{\label{figM_Bardeen4}\includegraphics[width=7.75cm]{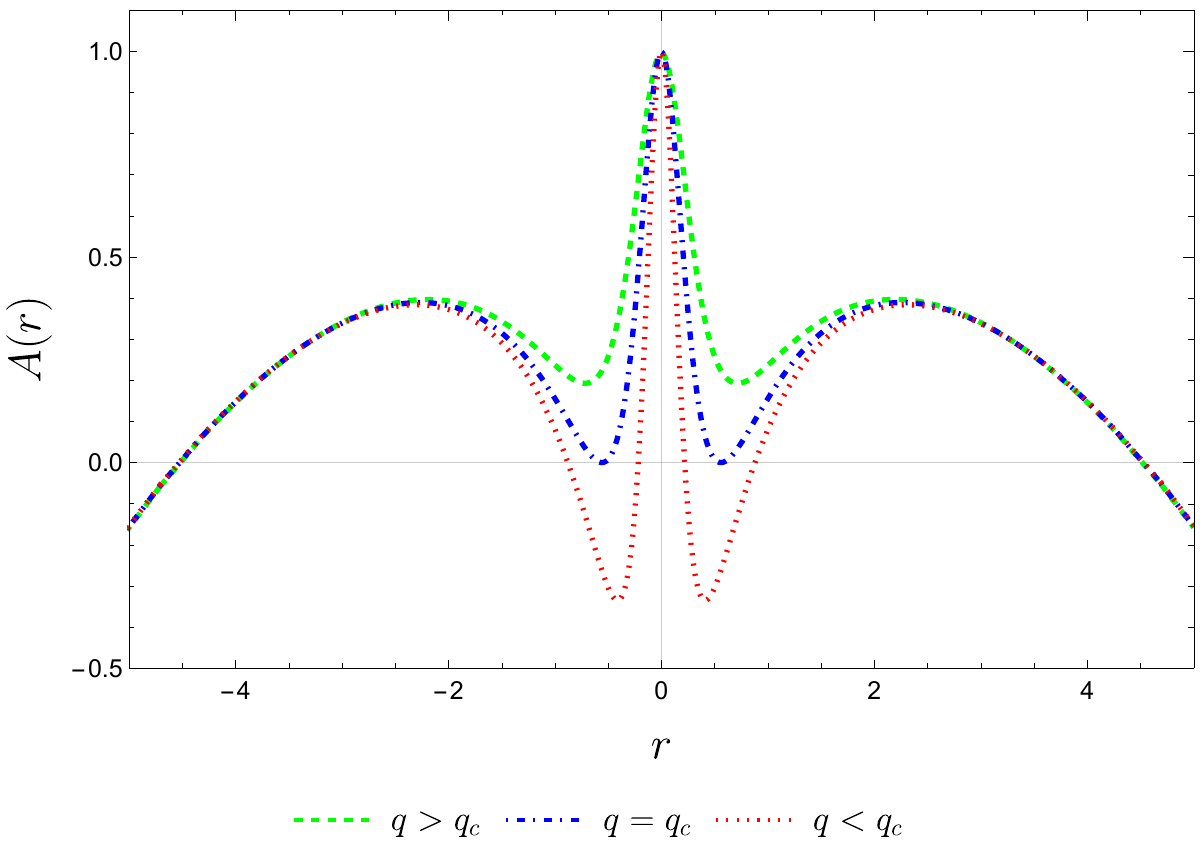}}
\subfigure[~Type-V solution: $\{q=0.75,M=2.0,\Lambda=-0.2\}$] 
{\label{figM_Bardeen5}\includegraphics[width=7.75cm]{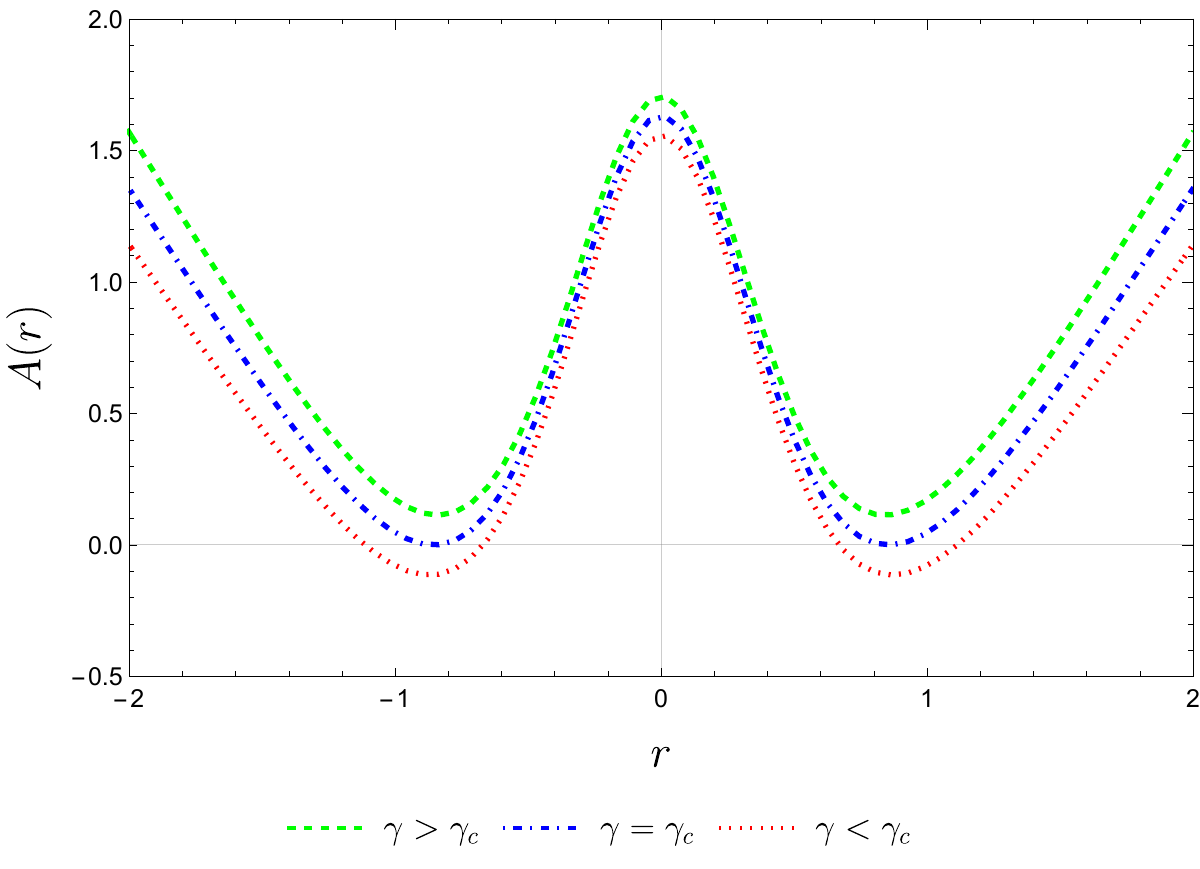}}
\hspace{0.75cm}
\subfigure[~Type-VI solution: $\{q=0.75,M=1.0,\Lambda=0.2 \}$] 
{\label{figM_Bardeen6}\includegraphics[width=7.75cm]{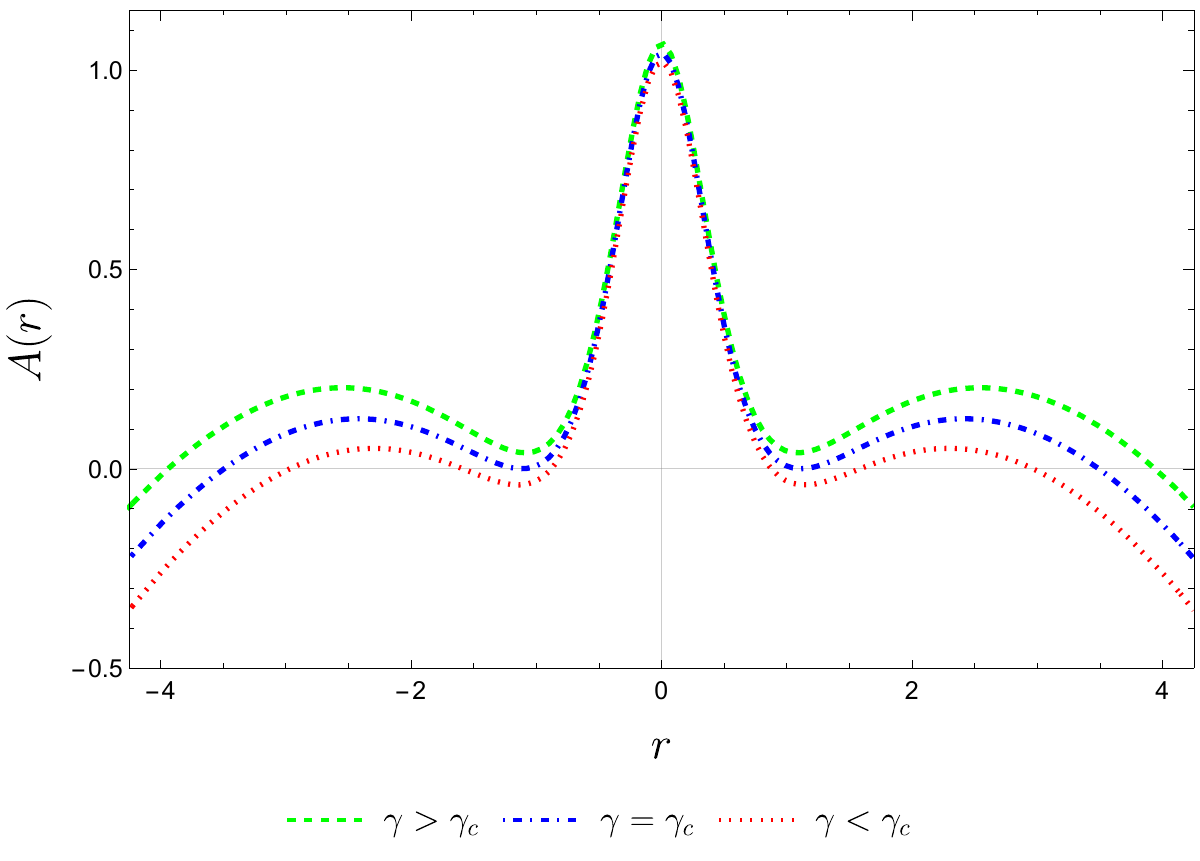}}
\caption{The plot depicts $A(r)$ for the six types of solutions described in Sec. \ref{BB_Bardeen} (see definitions in the main text), found by combining values for the four parameters $\{M,q,\Lambda,\gamma\}$ characterizing the solutions. See the text for more details.}
\label{figM_Bardeen}
\end{figure*}


{\it (i) Type-I solution: $\{q = 0.5, \Lambda = -0.2,\gamma=0.01\}$}

In this case the critical mass reads $M_c = 0.686$. The behaviour of the metric function~\eqref{a2_BB} is shown in Fig.~\ref{figM_Bardeen1} for three mass scenarios: $M > M_c$, $M = M_c$ and $M < M_c$. When the mass exceeds the critical mass, $M > M_c$, we observe a total of four horizons in the entire $r$ region, the innermost horizons being the Cauchy horizons and the outermost being the two event horizons.  At $r=0$ we have the throat of the wormhole, which is hidden behind these horizons. For the case where $M=M_c$, we have two degenerate horizons, one at $r > 0$ and the other at $r < 0$,  connected by the throat at $r=0$. In the case where $M < M_c$, we find the geometry of a traversable wormhole, a scenario similar to that in Fig. \ref{figM1_BB1} for $M < M_c$, Fig. \ref{figM1_BB3} for $q > q_c$ and Fig. \ref{figM1_BB5} for $\gamma>\gamma_c$. \\


{\it (ii) Type-II solution: $\{q=0.5,\Lambda=0.2,\gamma=0.01\}$}

In this case we find the critical mass value $M_c=0.811$. The scenarios $M > M_c$, $M = M_c$ and $M < M_c$ are illustrated in Fig. \ref{figM_Bardeen2}. When  $M > M_c$, two event horizons are formed, one in each $r$ region, connected at $r=0$ by a bounce in the radial function. When $M=M_c$, we observe the formation of four horizons, the innermost being Cauchy horizons, while the outermost ones are degenerate horizons that arise from the union of cosmological and event horizons. And when  $M < M_c$, we observe the formation of six horizons: the two innermost ones (on each side) are Cauchy horizons, the  two intermediate ones are the event horizons, and the two outermost ones are the cosmological horizons.\\


{\it (iii) Type-III solution: $\{\gamma=0.01,M=2.0,\Lambda=-0.2 \}$}

In this case we find the  critical charge value of $q_c=1.204$. The behavior of $A(r)$ is shown in Fig. \ref{figM_Bardeen3} for the cases $q > q_c$, $q=q_c$, and $q < q_c$. When $q > q_c$, we observe a similar behavior to the curve described in Fig. \ref{figM_Bardeen1} for the case where $M < M_c$, finding a traversable wormhole geometry. Similarly, when $q=q_c$ we find the same behaviour as that of $M=M_c$ in Fig. \ref{figM_Bardeen1}. Finally, when  $q < q_c$ the metric function exhibits the same behavior as the one of Fig. \ref{figM_Bardeen3} with $M > M_c$. \\


{\it (iv) Type-IV solution: $\{\gamma = 0.01,M = 0.5,\Lambda = 0.12\}$}

Now the critical charge reads $q_c = 0.389$. The behavior of $A(r)$ is shown in Fig. \ref{figM_Bardeen4} for the cases $q > q_c$, $q = q_c$, and $q < q_c$. When $q > q_c$ we observe a similar behavior as in Fig. \ref{figM_Bardeen2} for the case $q > q_c$; for $q=q_c$ we note the presence of degenerate horizons and two cosmological horizons, one each in the regions $r>0$ and $r<0$; and when  $q < q_c$, it shows the same behavior as in Fig. \ref{figM_Bardeen2}, but with $M < M_c$.\\


{\it (v) Type-V solution: $\{q=0.75,M=2.0,\Lambda=-0.2\}$}

Now we obtain the critical value of $\gamma_c=0.792$. The behavior of $A(r)$ is shown in Fig. \ref{figM_Bardeen5} for the scenarios $\gamma < \gamma_c$, $\gamma=\gamma_c$ and $\gamma>\gamma_c$. When $\gamma>\gamma_c$ we have a similar behavior to the cases of $M<M_c$ and $q>q_c$ as described in Fig. \ref{figM_Bardeen1} and \ref{figM_Bardeen3}; when $\gamma=\gamma_c$ a similar behaviour as Fig. \ref{figM_Bardeen1} for $M=M_c$ and Fig. \ref{figM_Bardeen3} for $q=q_c$; and when $\gamma<\gamma_c$ a similar situation as in Fig. \ref{figM_Bardeen1} when  $M > M_c$.\\


{\it (vi) Type-VI solution: $\{q=0.75,M=1.0,\Lambda=0.2 \}$}

Under these conditions we obtain $\gamma_c=0.107$. The behavior of $A(r)$ as a function of $r$ is shown in Fig. \ref{figM_Bardeen6} for the values $\gamma> \gamma_c$, $\gamma=\gamma_c$ and $ \gamma < \gamma_c$. When  $\gamma>\gamma_c$ we observe a similar behavior as in Fig. \ref{figM_Bardeen4} for the case of $q > q_c$; when $\gamma=\gamma_c$ we also have a similar scenario as in Fig. \ref{figM_Bardeen4} for $q= q_c$; and for $\gamma < \gamma_c$ a similar behaviour as is Fig. \ref{figM_Bardeen4} for $q<q_c$.

\subsection{Kretschmann scalar}

As for the regularity properties of this family of solutions, we compute the Kretschmann scalar as
\begin{widetext}
\begin{align}
    &K=	
\frac{1}{3\left(q^{2}+r^{2}\right)^{7}}\left\{ 12M^{2}\left(4q^{8}-44q^{6}r^{2}+169q^{4}r^{4}-68q^{2}r^{6}+12r^{8}\right)+4Mq^{2}\left(q^{2}+r^{2}\right)\right.	
	\nonumber\\
&+r^{6}\left(2\Lambda\sqrt{q^{2}+r^{2}}-30\gamma\right)-3q^{2}\left(16r^{2}\sqrt{q^{2}+r^{2}}+15\gamma r^{4}\right)+24r^{4}\sqrt{q^{2}+r^{2}}	\nonumber\\
& +q^{6}\left(4\Lambda\sqrt{q^{2}+r^{2}}-6\gamma\right) +q^{4}r^{2}\left(2\Lambda\sqrt{q^{2}+r^{2}}-21\gamma\right)	\nonumber
	\\
&+\left(q^{2}+r^{2}\right)^{3}\left[4\Lambda^{2}q^{8}+8r^{6}\left(3\gamma^{2}-3\gamma\Lambda\sqrt{q^{2}+r^{2}}+\Lambda^{2}r^{2}\right)+q^{6}\left(27\gamma^{2}-20\gamma\Lambda\sqrt{q^{2}+r^{2}}+16\Lambda\left(\Lambda r^{2}-1\right)\right)+\right.	\nonumber
	\\
&\left.\left.4q^{2}r^{4}\left(12\gamma^{2}-13\gamma\Lambda\sqrt{q^{2}+r^{2}}+2\Lambda\left(3\Lambda r^{2}-1\right)\right)+q^{4}\left(-48\gamma\sqrt{q^{2}+r^{2}}\left(\Lambda r^{2}-1\right)+51\gamma^{2}r^{2}+4\left(7\Lambda^{2}r^{4}-6\Lambda r^{2}+9\right)\right)\right]\right\} 	.
	\label{K_BB2}
\end{align}
\end{widetext}
which provides a finite value at $r\rightarrow0$ and, therefore, likewise in the previous case the geometry is everywhere regular.

\subsection{Matter Lagrangian reconstruction}

Using Eq. \eqref{a_BB}, we obtain the NLED Lagrangian density as
\begin{eqnarray} 
{\cal L}_{\rm NLED}(r) &=& f_{0}+\frac{2f_{1}q^{2}}{\sqrt{q^{2}+r^{2}}}+\frac{q^{2}}{21\kappa^{2}\left(q^{2}+r^{2}\right)^{7/2}}\times
\nonumber\\
&&\times\bigg[M\left(41q^{2}+231r^{2}\right)+7\left(q^{2}+r^{2}\right)\times\nonumber\\
&&\times\left(\gamma\left(q^{2}+r^{2}\right)-\sqrt{q^{2}+r^{2}}\right)\bigg],\label{L_BB2} 
\end{eqnarray}
its derivative as
\begin{eqnarray} 
{\cal L}_F(r)&=&\left(q^2+r^2\right)^{3/2} 
\Big\{f_1-\Big[190Mq^{2}-3\gamma\left(q^{2}+r^{2}\right)^{2}
	\nonumber\\
&&\hspace{-1.5cm}-165M\left(q^{2}+r^{2}\right)+4\left(q^{2}+r^{2}\right)^{3/2}\Big]\Big/\left[6\kappa^{2}\left(q^{2}+r^{2}\right)^{3}\right]\Big\} ,\label{LF_BB2} \nonumber \\
\end{eqnarray} 
and the scalar field potential as
%
\begin{eqnarray} 
&V(r)= \frac{2 q^2 }{3 \kappa ^2} \left(\frac{6 M \left(7 r^2-8 q^2\right)}{35 \left(q^2+r^2\right)^{7/2}}+\frac{\gamma }{\left(q^2+r^2\right)^{3/2}}-\frac{\Lambda }{q^2+r^2}\right),
\label{V2_BB20} 
\end{eqnarray}
respectively. Expressing $r(F)$ and $r(\cal{\varphi})$, we determine
\begin{align}
  & {\cal L} _{\rm NLED} (F)  = 	f_0+2 \sqrt[4]{2} f_1 \sqrt{F} q \sqrt{\frac{q}{\sqrt{F}}}-\frac{2 F}{3 \kappa ^2}
	\nonumber \\   
 & -\frac{380\ 2^{3/4} F^2 M \sqrt{\frac{q}{\sqrt{F}}}}{21 \kappa ^2}+\frac{2^{3/4} \gamma  F \sqrt{\frac{q}{\sqrt{F}}}}{3 \kappa ^2}
 \nonumber \\
& +\frac{22 \sqrt[4]{2} F^{3/2} M \sqrt{\frac{q}{\sqrt{F}}}}{\kappa ^2 q}.
 \label{L2_BB2} 
\end{align}  
and
\begin{eqnarray}
V(\varphi) &=& -\frac{2 \gamma  \cos ^4\left(\varphi \sqrt{\kappa ^2 (-\epsilon )}\right) \sqrt{q^2 \sec ^2\left(\varphi \sqrt{\kappa ^2 (-\epsilon )}\right)}}{3 \kappa ^2 q^2}
\nonumber\\
&& \hspace{-1.2cm} +\frac{4 M \sin ^2\left(\varphi \sqrt{\kappa ^2 (-\epsilon )}\right) \cos ^6\left(\varphi \sqrt{\kappa ^2 (-\epsilon )}\right) }{5 \kappa ^2 q^4}\times
	\nonumber \\
&& \hspace{-1.2cm}\times \sqrt{q^2 \sec ^2\left(\varphi \sqrt{\kappa ^2 (-\epsilon )}\right)} -\frac{2 \Lambda  \cos ^2\left(\varphi \sqrt{\kappa ^2 (-\epsilon )}\right)}{3 \kappa ^2}
\nonumber\\
&& \hspace{-1.2cm}  -\frac{32 M \cos ^8\left(\varphi \sqrt{\kappa ^2 (-\epsilon )}\right) \sqrt{q^2 \sec ^2\left(\varphi \sqrt{\kappa ^2 (-\epsilon )}\right)}}{35 \kappa ^2 q^4}.\label{V2_BB2}
\end{eqnarray} 

Considering again $f_0=0$, $f_1=0$, $\lambda=0$ and $\Lambda=0$ in the Lagrangian of equation~\eqref{L2_BB}, we obtain the same expression as the Lagrangian of GR, found in Ref.~\cite{Rodrigues2023}. The central limit, $F\gg1$, of the corresponding NLED Lagrangian reads as 
\begin{eqnarray}
    {\cal L} _{\rm NLED} (F)  \sim  f_{0}+2\sqrt[4]{2}f_{1}q^{3/2}\sqrt[4]{F}+\frac{22\sqrt[4]{2}F^{5/4}M}{\kappa^{2}\sqrt{q}}\nonumber\\
-\frac{2F}{3\kappa^{2}}   
   -\frac{380M2^{3/4}\sqrt{q}F^{7/4}}{21\kappa^{2}}+\frac{2^{3/4}\gamma\sqrt{q}F^{3/4}}{3\kappa^{2}}.\label{LF4_BB}
\end{eqnarray}
which is again a non-Maxwellian behaviour there.

\begin{figure}[t!]
\includegraphics[scale=0.45]{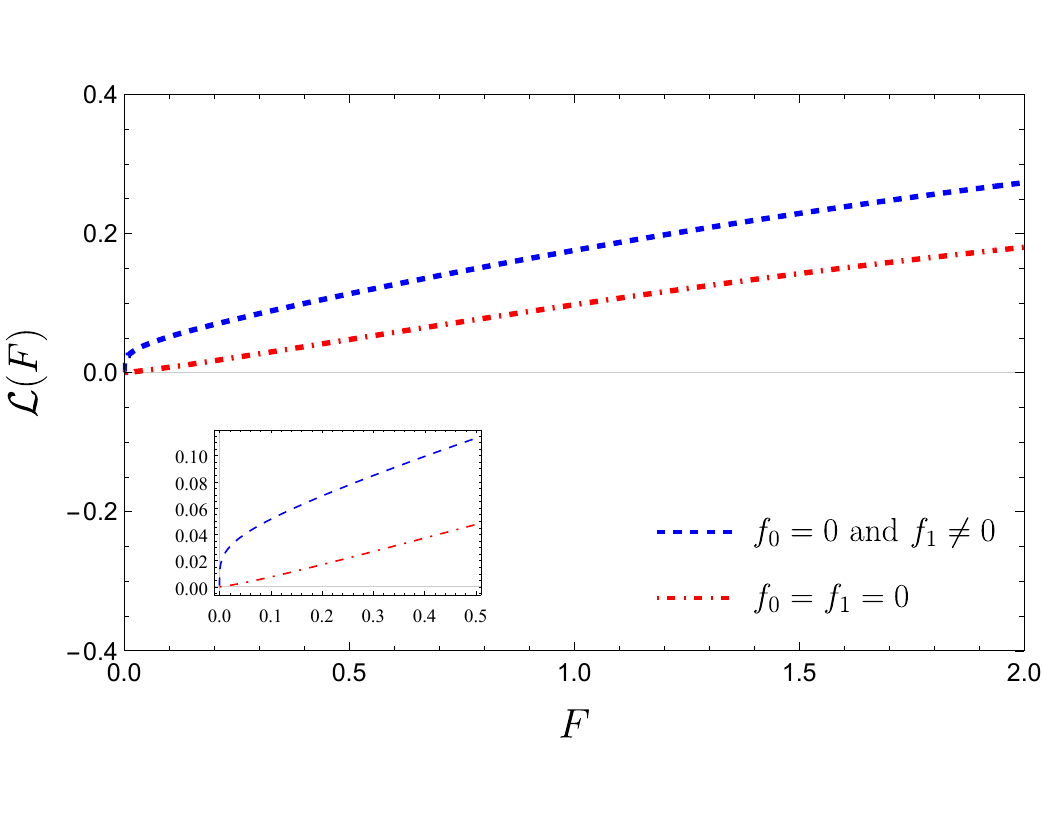}
\caption{The NLED Lagrangian ${\cal L}(F) $, given by Eq.~\eqref{L2_BB2}. The blue dashed curve represents the behavior of ${\cal L}(F)$ versus $F$ with $f_0=0$ and $f_1=0.2$, while the red dotted-dashed curve corresponds to $f_0=f_1=0$. We have taken the values of the constants as  $\{\gamma=0.1,\Lambda=-0.2,q=0.3,M=2.0 \}$.} 
\label{LxF_BB2}
\end{figure}

\begin{figure}[t!]
\includegraphics[scale=0.53]{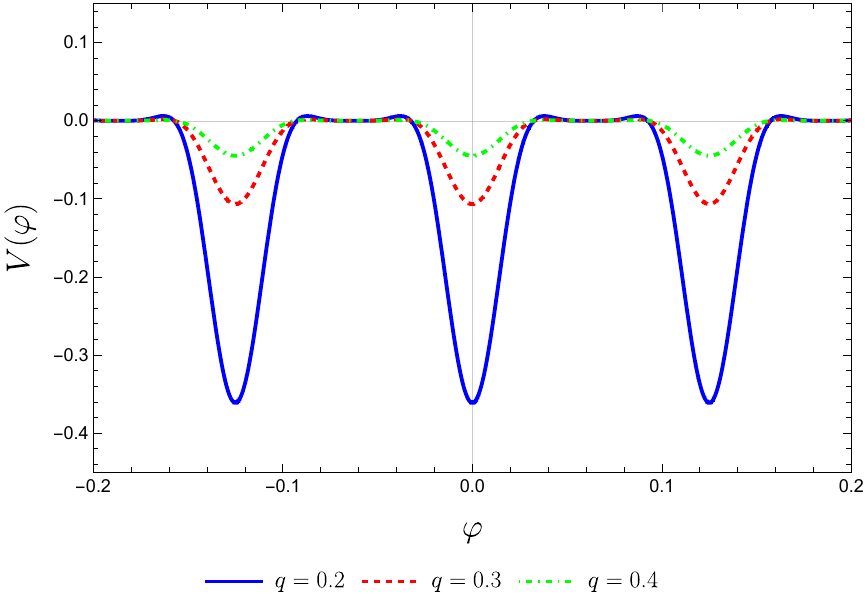}
\caption{ The scalar field potential $V(\varphi)$, described by Eq.~\eqref{V2_BB2} for different values of the charge taking $\epsilon=-1$,  for the values $\{\gamma=0.1,\Lambda=-0.2,M=2.0\}$.} 
\label{Vxphi_BB2}
\end{figure}

Fig. \ref{LxF_BB2} shows the behavior of the NLED Lagrangian density using Eq.~\eqref{L2_BB2} for the values $\{f_0=0,f_1=0.2\}$ (dashed blue curve), and $\{f_0=0,f_1=0\}$ (dotted-dashed red curve). Similarly, in Fig.~\ref{Vxphi_BB2} we depict the behaviour of the scalar field potential following  Eq. $\eqref{V2_BB2}$.

\section{Summary and Conclusion}\label{sec:concl}

In this paper we have studied spherically symmetric geometries implementing a bouncing in the radial function within a recently proposed theory of gravity known as Cotton Gravity (CG). Such geometries are inspired by a recently found vacuum solution of CG which generalizes the Schwarzschild solution of GR via a new parameter $\gamma$, and are further extended towards the black bounce mechanism to avoid the focusing of geodesics ascribed to the singularity theorems.  We have investigated such solutions by coupling the field equations of CG to a combination of a NLED and a scalar field. Solving the corresponding field equations we have found the free functions ${\cal L}_{\rm NLED }(r)$ and ${\cal L}_F(r)$ that characterize the NLED sector, and the scalar field potential $V(r)$ associated to a specific scalar field found within the context of GR. 

Subsequently, we have studied two generalizations of bouncing solutions found within GR, namely the  Simpson-Visser and Bardeen type solutions, both understood to be supported by magnetic NLED plus scalar fields. Such geometries differ from their GR counterparts  via an additional term depending on the parameter $\gamma$ of CG, and further supplemented by a cosmological constant term given the fact that the latter naturally arises as an integration constant in CG. Given the presence of four constants in our model, $\{M,q,\Lambda,\gamma\}$, we used an approach in which we set three of them and vary the remaining one (keeping always fixed the cosmological constant term $\Lambda$ to both positive and negative values), thus leading to six types of solutions on each case.

For the first model we used a generalized version of the bouncing metric of the Simpson-Visser type and simultaneously solved the equations \eqref{rH} and \eqref{der_a} to find the critical parameters for $\{M_c,q_c,\gamma_c\}$. For $\Lambda < 0$, in the case of the critical mass $M_c$,  when $M > M_c$ we found an event horizon on both sides of the bounce $r=0$, which for $M = M_c$ merge into a degenerate horizon at $r=0$, while if $M < M_c$ a traversable wormhole is formed. In the case of $\Lambda > 0$, no horizons are formed for $M > M_c$, while degenerate horizons are observed for $M = M_c$, and $M < M_c$, two horizons are observed on each side of the bounce, the outermost ones being cosmological horizons and the two innermost ones event horizons. On the other hand, for $\Lambda < 0$, in the case of critical charge, we observe that there is no horizon for $q > q_c$, when $q = q_c$ we have degenerate horizons, whereas for $q < q_c$ we observe two horizons. In the scenario where $\Lambda > 0$, we identify the presence of two cosmological horizons in all cases, whether $q > q_c$, $q = q_c$, or $q < q_c$.  We also analyzed the Kretschmann scalar of this model finding its regularity everywhere and, in particular, at the bounce location $r=0$. Finally, we verified via direct computation of their specific shapes that such bouncing geometries are supported by a specific combination of a NLED Lagrangian and a scalar field potential. Our setting recovers the GR one when we choose $f_0=f_1=\gamma=0$.

For the second model, we adopted Bardeen's regularized metric function and also determined the solutions for the horizons numerically. In the case of critical mass, for $\Lambda < 0$, when $M > M_c$ we observe four horizons, where the innermost horizons are Cauchy horizons on each side of the bounce; when $M = M_c$, degenerate horizons are observed; and if $M < M_c$ we find a traversable wormhole. For $\Lambda > 0$, if $M > M_c$ we observe two horizons, when $M = M_c$ we find  four horizons, the innermost of which are the Cauchy horizons. In the scenario where $M < M_c$, we discover six horizons, the innermost of which is the Cauchy horizon and the outermost being the cosmological horizons. For the critical charge in which $\Lambda < 0$, if $q > q_c$ there is no horizon, if $q = q_c$ we observe a degenerate horizons on each side of the bounce, while if $q < q_c$ we have two horizons on each side (Cauchy and event). For $\Lambda > 0$, we also observe that there are cosmological horizons in all scenarios ($q > q_c$, $q = q_c$, $q < q_c$) and, in particular, for $q < q_c$ a total of three horizons on each side are observed. With the critical parameter $\gamma$ for $\Lambda<0$, if $\gamma > \gamma_c$, we have a wormhole, for $\gamma =\gamma_c$, degenerate horizons occur, and if $\gamma < \gamma_c$ we have four horizons in total, with the innermost ones being Cauchy horizons. In the case where $\Lambda > 0$, two horizons are formed for $\gamma > \gamma_c$, degenerate horizons are formed for $\gamma = \gamma_c$, and for $\gamma < \gamma_c$ we have six horizons. Similarly as in the Simpson-Visser geometry we verified the regularity of the  Kretschmann scalar and found the explicit shapes of the NLED Lagrangian density and the scalar potential.

The bottom line of our analysis is that CG naturally allows for suitable generalizations of bouncing geometries of GR in terms of the new CG parameter $\gamma$, extracted out of the class of vacuum solutions found within this theory. The geometries built this way display promising features to be studied within the context of black hole thermodynamics, the analysis 
 the stability of solutions through the analysis of perturbations, gravitational lensing and black hole shadows, topics in which we hope to report soon.


\acknowledgments{
MER thanks Conselho Nacional de Desenvolvimento Cient\'ifico e Tecnol\'ogico - CNPq, Brazil, for partial financial support. This study was financed in part by the Coordena\c{c}\~{a}o de Aperfei\c{c}oamento de Pessoal de N\'{i}vel Superior - Brasil (CAPES) - Finance Code 001.
FSNL acknowledges support from the Funda\c{c}\~{a}o para a Ci\^{e}ncia e a Tecnologia (FCT) Scientific Employment Stimulus contract with reference CEECINST/00032/2018, and funding through the research grants UIDB/04434/2020, UIDP/04434/2020 and PTDC/FIS-AST/0054/2021.
DRG is supported by the Spanish Agencia Estatal de Investigación Grant No. PID2022-138607NB-I00, funded by MCIN/AEI/10.13039/501100011033, UE, and ERDF A way of making Europe.}




\begin{thebibliography}{99}

\bibitem{Harada:2021bte}
J.~Harada,
``Emergence of the Cotton tensor for describing gravity,''
Phys. Rev. D \textbf{103} (2021) no.12, L121502
[arXiv:2105.09304 [gr-qc]].

\bibitem{Harada}
J.~Harada,
``Gravity at cosmological distances: Explaining the accelerating expansion without dark energy,''
Phys. Rev. D \textbf{108} (2023) no.4, 044031
[arXiv:2308.02115 [gr-qc]].

\bibitem{Cottonpaper} 
E. Cotton, Sur les variétés à trois dimensions, Ann. Fac. Sci. Toulouse 2. 1, 385 (1899).

\bibitem{Jackiw:2004qm}
R.~Jackiw,
``A pure cotton kink in a funny place,'' in 10th International Symposium on Particles, Strings and Cos-
mology (PASCOS 04 and Pran Nath Fest) (2004) pp. 507–515.

\bibitem{Deser:2004wd}
S.~Deser, R.~Jackiw and S.~Y.~Pi,
``Cotton blend gravity pp waves,''
Acta Phys. Polon. B \textbf{36} (2005), 27-34
[arXiv:gr-qc/0409011 [gr-qc]].



\bibitem{Harada:2022edl}
J.~Harada,
``Cotton gravity and 84 galaxy rotation curves,''
Phys. Rev. D \textbf{106} (2022) no.6, 064044
[arXiv:2209.04055 [gr-qc]].

\bibitem{Sussman:2023eep}
R.~A.~Sussman and S.~Najera,
``Exact solutions of Cotton Gravity,''
[arXiv:2312.02115 [gr-qc]].

\bibitem{Mantica:2022flg}
C.~A.~Mantica and L.~G.~Molinari,
``Codazzi tensors and their space-times and Cotton gravity,''
Gen. Rel. Grav. \textbf{55} (2023) no.4, 62
[arXiv:2210.06173 [gr-qc]].


\bibitem{Sussman:2023wiw}
R.~A.~Sussman and S.~Najera,
``Cotton Gravity: the cosmological constant as spatial curvature,''
[arXiv:2311.06744 [gr-qc]].


\bibitem{Gogberashvili:2023wed}
M.~Gogberashvili and A.~Girgvliani,
``General spherically symmetric solution of Cotton gravity,''
Class. Quant. Grav. \textbf{41} (2024) no.2, 025010
[arXiv:2308.03342 [gr-qc]].

\bibitem{Mantica:2023ssd}
C.~A.~Mantica and L.~G.~Molinari,
``Friedmann equations in the Codazzi parametrization of Cotton and extended theories of gravity and the dark sector,''
Phys. Rev. D \textbf{109} (2024) no.4, 044059
[arXiv:2312.02784 [gr-qc]].

\bibitem{Xia:2024tps}
P.~Xia, D.~Zhang, X.~Ren, B.~Wang and Y.~C.~Ong,
``Constrain the linear scalar perturbation theory of Cotton gravity,''
[arXiv:2405.07209 [astro-ph.CO]].

\bibitem{Mo:2024rfq}
G.~Mo, Q.~Wang, X.~Ren, W.~Yan, Y.~C.~Ong and W.~Luo,
``Testing Cotton gravity as dark matter substitute with weak lensing,''
[arXiv:2405.07215 [astro-ph.CO]].

\bibitem{Junior:2023ixh}
J.~T.~S.~S.~Junior, F.~S.~N.~Lobo and M.~E.~Rodrigues,
``(Regular) Black holes in conformal Killing gravity coupled to nonlinear electrodynamics and scalar fields,''
Class. Quant. Grav. \textbf{41} (2024) no.5, 055012
[arXiv:2310.19508 [gr-qc]].

\bibitem{Junior:2024vrv}
J.~T.~S.~S.~Junior, F.~S.~N.~Lobo and M.~E.~Rodrigues,
``Black bounces in conformal Killing gravity,''
Eur. Phys. J. C \textbf{84} (2024) no.6, 557
[arXiv:2405.09702 [gr-qc]].


\bibitem{Mantica:2023stl}
C.~A.~Mantica and L.~G.~Molinari,
``Note on Harada\textquoteright{}s conformal Killing gravity,''
Phys. Rev. D \textbf{108} (2023) no.12, 124029
[arXiv:2308.06803 [gr-qc]].
 

\bibitem{Harada:2023afu}
J.~Harada,
``Dark energy in conformal Killing gravity,''
Phys. Rev. D \textbf{108} (2023) no.10, 104037
[arXiv:2308.07634 [gr-qc]].



\bibitem{Barnes:2023uru}
A.~Barnes,
``Vacuum Static Spherically Symmetric Spacetimes in Harada's Theory,''
[arXiv:2309.05336 [gr-qc]].

\bibitem{Barnes:2023qfi}
A.~Barnes,
``Harada-Maxwell Static Spherically Symmetric Spacetimes,''
[arXiv:2311.09171 [gr-qc]].

\bibitem{Barnes:2024vjq}
A.~Barnes,
``pp-waves in conformal Killing gravity,''
[arXiv:2404.09310 [gr-qc]].

\bibitem{Mantica:2024mun}
C.~A.~Mantica and L.~G.~Molinari,
``Conformal Killing cosmology -- Geometry, dark sector, growth of structures,''
[arXiv:2404.11468 [gr-qc]].




\bibitem{Clement:2023tyx}
G.~Cl\'ement and K.~Nouicer,
``Cotton gravity is not predictive,''
[arXiv:2312.17662 [gr-qc]].


\bibitem{Sussman:2024iwk}
R.~A.~Sussman, C.~A.~Mantica, L.~G.~Molinari and S.~N\'ajera,
``Response to a critique of ''Cotton Gravity'',''
[arXiv:2401.10479 [gr-qc]].

\bibitem{Clement:2024pjl}
G.~Cl\'ement and K.~Nouicer,
``Farewell to Cotton gravity,''
[arXiv:2401.16008 [gr-qc]].

\bibitem{Sussman:2024qsg}
R.~A.~Sussman, C.~A.~Mantica, L.~G.~Molinari and S.~N\'ajera,
``Second Response to the critique of ''Cotton Gravity'',''
[arXiv:2402.01992 [gr-qc]].


\bibitem{Senovilla:2014gza}
J.~M.~M.~Senovilla and D.~Garfinkle,
``The 1965 Penrose singularity theorem,''
Class. Quant. Grav. \textbf{32} (2015) no.12, 124008
[arXiv:1410.5226 [gr-qc]].

\bibitem{Bardeen}
J.M. Bardeen, Non-singular general-relativistic gravitational collapse, in Proceedings of of International Conference GR5,Tbilisi, USSR (1968), p. 174


\bibitem{Ayon-Beato:2000mjt}
E.~Ayon-Beato and A.~Garcia,
``The Bardeen model as a nonlinear magnetic monopole,''
Phys. Lett. B \textbf{493} (2000), 149-152
[arXiv:gr-qc/0009077 [gr-qc]].


\bibitem{Simpson:2018tsi}
A.~Simpson and M.~Visser,
``Black-bounce to traversable wormhole,''
JCAP \textbf{02} (2019), 042
[arXiv:1812.07114 [gr-qc]].


\bibitem{Lobo:2020kxn}
F.~S.~N.~Lobo, A.~Simpson and M.~Visser,
``Dynamic thin-shell black-bounce traversable wormholes,''
Phys. Rev. D \textbf{101} (2020) no.12, 124035
[arXiv:2003.09419 [gr-qc]].

\bibitem{Junior:2022zxo}
E.~L.~B.~Junior and M.~E.~Rodrigues,
``Black-bounce in f(T) gravity,''
Gen. Rel. Grav. \textbf{55} (2023) no.1, 8
[arXiv:2203.03629 [gr-qc]].


\bibitem{Junior:2023qaq}
J.~T.~S.~S.~Junior and M.~E.~Rodrigues,
``Coincident $f(\mathbb {Q})$ gravity: black holes, regular black holes, and black bounces,''
Eur. Phys. J. C \textbf{83} (2023) no.6, 475
[arXiv:2306.04661 [gr-qc]].

\bibitem{Junior:2024xmm}
J.~T.~S.~S.~Junior, F.~S.~N.~Lobo and M.~E.~Rodrigues,
``Black holes and regular black holes in coincident $f({\mathbb {Q}},{\mathbb {B}}_Q)$ gravity coupled to nonlinear electrodynamics,''
Eur. Phys. J. C \textbf{84} (2024) no.3, 332
[arXiv:2402.02534 [gr-qc]].


\bibitem{Rodrigues:2022mdm}
M.~E.~Rodrigues and M.~V.~d.~Silva,
``Black-bounces with multiple throats and anti-throats,''
Class. Quant. Grav. \textbf{40} (2023) no.22, 225011
[arXiv:2204.11851 [gr-qc]].


\bibitem{Huang:2019arj}
H.~Huang and J.~Yang,
``Charged Ellis Wormhole and Black Bounce,''
Phys. Rev. D \textbf{100} (2019) no.12, 124063
[arXiv:1909.04603 [gr-qc]].


\bibitem{Lobo:2020ffi}
F.~S.~N.~Lobo, M.~E.~Rodrigues, M.~V.~de Sousa Silva, A.~Simpson and M.~Visser,
``Novel black-bounce spacetimes: wormholes, regularity, energy conditions, and causal structure,''
Phys. Rev. D \textbf{103} (2021) no.8, 084052
[arXiv:2009.12057 [gr-qc]].


\bibitem{Nascimento:2020ime}
J.~R.~Nascimento, A.~Y.~Petrov, P.~J.~Porfirio and A.~R.~Soares,
``Gravitational lensing in black-bounce spacetimes,''
Phys. Rev. D \textbf{102} (2020) no.4, 044021
[arXiv:2005.13096 [gr-qc]].


\bibitem{Tsukamoto:2020bjm}
N.~Tsukamoto,
``Gravitational lensing in the Simpson-Visser black-bounce spacetime in a strong deflection limit,''
Phys. Rev. D \textbf{103} (2021) no.2, 024033
[arXiv:2011.03932 [gr-qc]].


\bibitem{Cheng:2021hoc}
X.~T.~Cheng and Y.~Xie,
``Probing a black-bounce, traversable wormhole with weak deflection gravitational lensing,''
Phys. Rev. D \textbf{103}, no.6, 064040 (2021).


\bibitem{Tsukamoto:2021caq}
``Gravitational lensing by two photon spheres in a black-bounce spacetime in strong deflection limits,''
Phys. Rev. D \textbf{104} (2021) no.6, 064022
[arXiv:2105.14336 [gr-qc]].


\bibitem{Zhang:2022nnj}
J.~Zhang and Y.~Xie,
``Gravitational lensing by a black-bounce-Reissner\textendash{}Nordstr\"om spacetime,''
Eur. Phys. J. C \textbf{82} (2022) no.5, 471.


\bibitem{Guerrero:2021ues}
M.~Guerrero, G.~J.~Olmo, D.~Rubiera-Garcia and D.~S.~C.~G\'omez,
``Shadows and optical appearance of black bounces illuminated by a thin accretion disk,''
JCAP \textbf{08} (2021), 036
[arXiv:2105.15073 [gr-qc]].


\bibitem{Jafarzade:2021umv}
K.~Jafarzade, M.~Kord Zangeneh and F.~S.~N.~Lobo,
``Observational optical constraints of regular black holes,''
Annals Phys. \textbf{446} (2022), 169126
[arXiv:2106.13893 [gr-qc]].

\bibitem{Jafarzade:2020ova}
K.~Jafarzade, M.~Kord Zangeneh and F.~S.~N.~Lobo,
``Shadow, deflection angle and quasinormal modes of Born-Infeld charged black holes,''
JCAP \textbf{04} (2021), 008
[arXiv:2010.05755 [gr-qc]].

\bibitem{Jafarzade:2020ilt}
K.~Jafarzade, M.~Kord Zangeneh and F.~S.~N.~Lobo,
``Optical Features of AdS Black Holes in the Novel 4D Einstein-Gauss-Bonnet Gravity Coupled to Nonlinear Electrodynamics,''
Universe \textbf{8} (2022) no.3, 182
[arXiv:2009.12988 [gr-qc]].


\bibitem{Yang:2021cvh}
Y.~Yang, D.~Liu, Z.~Xu, Y.~Xing, S.~Wu and Z.~W.~Long,
``Echoes of novel black-bounce spacetimes,''
Phys. Rev. D \textbf{104} (2021) no.10, 104021
[arXiv:2107.06554 [gr-qc]].


\bibitem{Bambhaniya:2021ugr}
P.~Bambhaniya, S.~K, K.~Jusufi and P.~S.~Joshi,
``Thin accretion disk in the Simpson-Visser black-bounce and wormhole spacetimes,''
Phys. Rev. D \textbf{105} (2022) no.2, 023021
[arXiv:2109.15054 [gr-qc]].


\bibitem{Ou:2021efv}
M.~Y.~Ou, M.~Y.~Lai and H.~Huang,
``Echoes from asymmetric wormholes and black bounce,''
Eur. Phys. J. C \textbf{82} (2022) no.5, 452
[arXiv:2111.13890 [gr-qc]].


\bibitem{Guo:2021wid}
Y.~Guo and Y.~G.~Miao,
``Charged black-bounce spacetimes: Photon rings, shadows and observational appearances,''
Nucl. Phys. B \textbf{983} (2022), 115938
[arXiv:2112.01747 [gr-qc]].


\bibitem{Wu:2022eiv}
S.~R.~Wu, B.~Q.~Wang, D.~Liu and Z.~W.~Long,
``Echoes of charged black-bounce spacetimes,''
Eur. Phys. J. C \textbf{82} (2022) no.11, 998
[arXiv:2201.08415 [gr-qc]].


\bibitem{Tsukamoto:2022vkt}
N.~Tsukamoto,
``Retrolensing by two photon spheres of a black-bounce spacetime,''
Phys. Rev. D \textbf{105} (2022) no.8, 084036
[arXiv:2202.09641 [gr-qc]].


\bibitem{Mazza:2021rgq}
J.~Mazza, E.~Franzin and S.~Liberati,
``A novel family of rotating black hole mimickers,''
JCAP \textbf{04} (2021), 082
[arXiv:2102.01105 [gr-qc]].


\bibitem{Xu:2021lff}
Z.~Xu and M.~Tang,
``Rotating spacetime: black-bounces and quantum deformed black hole,''
Eur. Phys. J. C \textbf{81} (2021) no.10, 863
[arXiv:2109.13813 [gr-qc]].



\bibitem{Bronnikov:2022bud}
K.~A.~Bronnikov,
``Black bounces, wormholes, and partly phantom scalar fields,''
Phys. Rev. D \textbf{106} (2022) no.6, 064029
[arXiv:2206.09227 [gr-qc]].


\bibitem{Canate:2022gpy}
P.~Ca\~nate,
``Black bounces as magnetically charged phantom regular black holes in Einstein-nonlinear electrodynamics gravity coupled to a self-interacting scalar field,''
Phys. Rev. D \textbf{106} (2022) no.2, 024031
[arXiv:2202.02303 [gr-qc]].


\bibitem{Rodrigues2023}
M.~E.~Rodrigues and M.~V.~d.~S.~Silva,
``Source of black bounces in general relativity,''
Phys. Rev. D \textbf{107} (2023) no.4, 044064
[arXiv:2302.10772 [gr-qc]].



\bibitem{Pereira:2023lck}
C.~F.~S.~Pereira, D.~C.~Rodrigues, J.~C.~Fabris and M.~E.~Rodrigues,
``Black-bounce solution in k-essence theories,''
Phys. Rev. D \textbf{109} (2024) no.4, 044011
[arXiv:2309.10963 [gr-qc]].

\bibitem{Grumiller:2010bz}
D.~Grumiller,
``Model for gravity at large distances,''
Phys. Rev. Lett. \textbf{105} (2010), 211303
[erratum: Phys. Rev. Lett. \textbf{106} (2011), 039901]
[arXiv:1011.3625 [astro-ph.CO]].

\bibitem{Ellis:1973yv}
H.~G.~Ellis,
``Ether flow through a drainhole - a particle model in general relativity,''
J. Math. Phys. \textbf{14} (1973), 104-118.

\bibitem{Bronnikov:2021uta}
K.~A.~Bronnikov and R.~K.~Walia,
``Field sources for Simpson-Visser spacetimes,''
Phys. Rev. D \textbf{105} (2022) no.4, 044039
[arXiv:2112.13198 [gr-qc]].


\bibitem{Ansoldi:2008jw}
S.~Ansoldi,
``Spherical black holes with regular center: A Review of existing models including a recent realization with Gaussian sources,''
[arXiv:0802.0330 [gr-qc]].


\end{thebibliography}
\end{document}